\documentclass[twocolumn]{aastex61}
\usepackage{graphicx}
\usepackage{float}
\usepackage{longtable}
\usepackage{pgfplots}

\begin{document}
\title{Analytic model for tangential YORP}
\correspondingauthor{Oleksiy Golubov}
\email{oleksiy.golubov@karazin.ua}
\author[0000-0002-5720-2282]{Oleksiy Golubov}
\affiliation{Department of Aerospace Engineering Sciences, University of Colorado at Boulder \\
429 UCB, Boulder, CO, 80309, USA}
\affiliation{Karazin Kharkiv National University, 4 Svobody Sq., Kharkiv, 61022, Ukraine}
\affiliation{Institute of Astronomy of V. N. Karazin Kharkiv National University, 35 Sumska Str., Kharkiv, 61022, Ukraine}

\begin{abstract}
The tangential YORP effect (TYORP) plays a significant role in the dynamical evolution of asteroids, and up to now has only been studied numerically.
This paper describes the first analytic model of the TYORP effect.
Although the model rests on numerous physical and mathematical simplifications,
the final analytic expression for TYORP is found to be in agreement 
with the results of rigorous numeric simulations to the accuracy of several tens per cent.

The obtained analytic expression is used to estimate the TYORP produced by the non-flat surface of regolith, -- a contribution to TYORP that has never been considered. It is found that the contribution to TYORP arising from regolith can be comparable to the conventional TYORP produced by boulders.

Then, the analytic expression is fitted with a log-normal function
and used to integrate TYORP over all boulder sizes.
The general trend of TYORP for multiple boulders appears qualitatively similar to the trend of one boulder,
and also demonstrates a maximal TYORP at some particular rotation rate.
The obtained expression for integrated TYORP may be instrumental for simulations of evolution of asteroids subject to TYORP.

To conclude, the physical origin of TYORP is discussed in light of the constructed analytic model.

\end{abstract}


\section{Introduction}
The tangential YORP effect, or TYORP, appears when stones on the surface of an asteroid emit different amounts of infrared light eastward and westward, thus experiencing a net recoil force tangential to the asteroid's surface.
Until now, this effect has only been studied in numeric simulations \citep{golubov12,golubov14,sevecek15,sevecek16}.
Although it was generally understood that the effect was due to the non-linearity of the heat emission law,
the detailed physics of the effect remained obscure.
Moreover, the question remained whether the whole effect could be attributed to numeric artifacts.

In this article I propose a minimalistic analytic model of the effect, which is based on the following simplifying assumptions:

1. Instead of solving partial differential equation for the heat conduction in a boulder, the boulder is split into two parts,
the eastern part and the western part, and the mean temperature of each part is introduced. Then the mathematical description of the model boils down to a system of two ordinary differential equations for the mean temperatures.

2. The incoming solar energy as a function of time is approximated by the sum of its zeroth and first order Fourier terms; 
all higher-order terms are neglected.

3. The first order Fourier term is treated perturbatively, as if it were small compared to the zeroth order term.

After these simplifications the problem can be easily solved analytically,
and the TYORP drag can be calculated.
The result fits the numeric simulations surprisingly well.

In Section \ref{sec-theory}, I provide a derivation of the analytic expression for TYORP.
In Section \ref{sec-shapes}, the derived analytic expression is applied to different geometries, to test it and to make some new predictions.
In Section \ref{sec-multiple}, I simplify the analytic expression and integrate it over different boulder sizes, to evaluate the total TYORP experienced by an asteroid.
In Section \ref{sec-discussion}, I discuss how the derived analytic expression helps to better understand the physics of TYORP.	

\section{General theory\label{sec-theory}}
\subsection{Derivation of the heat conduction equations}
\begin{figure}
\centering
\begin{tikzpicture}

\draw [color=gray!20,fill=gray!20] (-6,0)--(-2,0)--(-2,-1.5)--(-6,-1.5)--cycle;
\draw[thick] (-6,0) -- (-2,0);

\begin{scope}
\path[clip] (-6,1.5)--(-2,1.5)--(-2,-1.5)--(-6,-1.5)--cycle;
\draw[thick] (-4,0) ellipse (1.5 and 1);
\fill[gray!50] (-4,0) ellipse (1.5 and 1);
\fill[gray!50] (-4.5,-1) arc(270:90:1) -- (-4.5,-1) --(-4,-1)--(-4,1)--(-4.5,1)--cycle;
\end{scope}

\draw[color=red,ultra thick] (-5.5,0) arc (180:90:1) ;
\draw[color=red,ultra thick] (-4.5,1)--(-4,1);
\draw[color=orange,ultra thick] (-5.5,0) arc (180:270:1) ;
\draw[color=orange,ultra thick] (-4.5,-1)--(-4,-1);
\draw[color=purple,ultra thick] (-4,-1) -- (-4,1);

\node[color=red] at (-5.4,1) {$S_\mathrm{sp}$};
\node[color=orange] at (-5.3,-1) {$S_\mathrm{reg}$};
\node[color=purple] at (-3.7,0.5) {$S_\mathrm{st}$};
\node[color=red] at (-4.7,0) {$V$};
\node[] at (-3.5,-0.5) {stone};
\node[] at (-3.9,-1.3) {regolith};
\node[] at (-3.5,1.2) {space};

\draw [color=gray!20,fill=gray!20] (-1.5,0)--(1.5,0)--(1.5,-1.5)--(-1.5,-1.5)--cycle;
\draw[thick] (-1.5,0) -- (1.5,0);

\fill[red!30] (0,-1) arc(270:90:1) -- (0,-1) --cycle;
\fill[blue!30] (0,-1) arc(-90:90:1) -- (0,-1) --cycle;

\draw[color=purple,ultra thick] (0,-1) -- (0,1);
\draw[color=blue,ultra thick] (1,0) arc (0:90:1) ;
\draw[color=cyan,ultra thick] (1,0) arc (0:-90:1) ;
\draw[color=red,ultra thick] (-1,0) arc (180:90:1) ;
\draw[color=orange,ultra thick] (-1,0) arc (180:270:1) ;

\node[color=red] at (-1,1) {$s_\mathrm{sp\ w}$};
\node[color=orange] at (-1,-1) {$s_\mathrm{reg\ w}$};
\node[color=blue] at (1,1) {$s_\mathrm{sp\ e}$};
\node[color=cyan] at (1,-1) {$s_\mathrm{reg\ e}$};
\node[color=purple] at (0.3,0.5) {$s_\mathrm{st}$};
\node[color=red] at (-0.5,0) {$v_\mathrm{w}$};
\node[color=blue] at (0.5,0) {$v_\mathrm{e}$};

\end{tikzpicture}
\caption{Illustration of volumes and surface areas in the heat conduction equations. \textit{Left:} Dimensional volume and surface areas in Eqn. (\ref{dimensional_heat_balance}). \textit{Right:} Dimensionless volumes and surface areas used for derivation of Eqn. (\ref{heat_balance2}).}
\label{fig:vs}%
\end{figure}
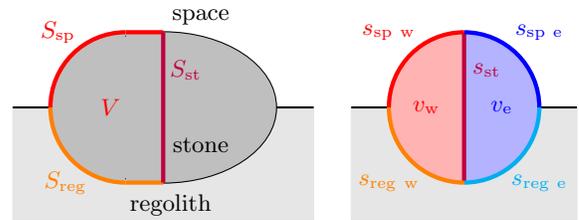

Heat balance within any volume part $V$ of a boulder is governed by the following heat conduction equation in the integral form,
\begin{eqnarray}
 C\rho\int\limits_V\frac{\partial T}{\partial t}\mathrm{d}V=
 \kappa\int\limits_{S_\mathrm{st}}\frac{\partial T}{\partial X_i}\mathrm{d}S_i+ \nonumber\\ 
 +(1-A)\int\limits_{S_\mathrm{sp}}I_i\mathrm{d}S_i-
 \epsilon\sigma \int\limits_{S_\mathrm{sp}}T^4\mathrm{d}S\,.
 \label{dimensional_heat_balance}
\end{eqnarray}
Here $T$ stands for the temperature.
The left-hand side describes the total heat energy increase in the volume $V$,
while the right-hand side is the sum of the heat conduction into this volume,
the direct solar heat absorbed by its open surface, 
and the negative heat emitted by the open surface.
The surface areas $S_\mathrm{st}$, $S_\mathrm{sp}$, and $S_\mathrm{reg}$ are parts of the volume's boundary bordering respectively
stone, space, and regolith, so that $S_\mathrm{st}+S_\mathrm{sp}+S_\mathrm{reg}=\partial V$ is the full boundary of the volume $V$ 
(see the left-hand panel of Figure \ref{fig:vs}).
The heat conductivity of the stone is $\kappa$, 
its heat capacity is $C$, the density is $\rho$, 
the hemispherical albedo is $A$, and the emissivity is $\epsilon$.
The heat conductivity of the regolith is assumed to be zero.
$\sigma$ is Stefan--Boltzmann's constant, and $\mathbf{I}$ is the vector of the incoming solar energy flux.

Now I nondimensionalize the variables.
Instead of time $t$, I use the rotation phase $\phi=\omega t$, with $\omega$ being the angular velocity of the asteroid.
By definition, $\phi=0$ at noon.
The characteristic scales of length and temperature are
\begin{equation}
 L_\mathrm{cond}=\frac{\kappa}{\left((1-A)\Phi\right)^{3/4}\left(\epsilon \sigma \right)^{1/4}}\,,
\end{equation}
\begin{equation}
 T_0=\sqrt[4]{\frac{(1-A)\Phi}{\epsilon\sigma}}\,,
\end{equation}
with $\Phi$ being the solar constant.
Here $T_0$ is the equilibrium temperature at the subsolar point, while $L_\mathrm{cond}$ is the distance at which the temperature difference $T_0$ creates heat flux equal to $A\Phi$.
I use these two scales to introduce the dimensionless variables $x_i=X_i/L_\mathrm{cond}$ and $\tau=T/T_0$.
The relative importance of heat conduction with respect to heat absorption and emission is characterized by the thermal parameter
\begin{equation}
 \theta=\frac{\left(C\rho\kappa\omega\right)^{1/2}}{\left((1-A)\Phi\right)^{3/4}\left(\epsilon \sigma \right)^{1/4}}.
\end{equation}

With these definitions, Eqn. (\ref{dimensional_heat_balance}) transforms into
\begin{eqnarray}
 \theta^2\int\limits_v\frac{\partial \tau}{\partial \phi}\mathrm{d}v=
 \int\limits_{s_\mathrm{st}}\frac{\partial \tau}{\partial x_i}\mathrm{d}s_i+
 \int\limits_{s_\mathrm{sp}}\frac{I_i}{\Phi}\mathrm{d}s_i-
 \int\limits_{s_\mathrm{sp}}\tau^4\mathrm{d}s\,.
 \label{heat_balance1}
\end{eqnarray}
Here $v$ and $s$ correspond to the same volumes and areas as before,
but measured in the dimensionless variables $x_i$
instead of the dimensional variables $X_i$.

Now I separate the boulder into the western and the eastern parts and apply Eqn. (\ref{dimensional_heat_balance}) to each part separately.
I assume that the boulder is symmetric, with the western and the eastern parts being mirror reflections of each other.
Let $v$ henceforth denote the dimensionless volume of each half of the boulder (either $v_\mathrm{w}$ or $v_\mathrm{e}$, see the right-hand panel of Figure \ref{fig:vs}),
and $s_\mathrm{sp}$ denote their equal dimensionless surface areas bordering space (either $s_\mathrm{sp\ e}$ or $s_\mathrm{sp\ w}$ in the right-hand panel of Figure \ref{fig:vs}).
I denote the mean dimensionless temperatures of the western and the eastern parts of the boulder via $\tau_\mathrm{w}$ and $\tau_\mathrm{e}$ correspondingly,
\begin{eqnarray}
\tau_\mathrm{w}&=& \frac{1}{v}\int\limits_{v_\mathrm{w}} \tau \, dv \,, \nonumber \\
\tau_\mathrm{e}&=& \frac{1}{v}\int\limits_{v_\mathrm{e}} \tau \, dv\,.
\end{eqnarray}
The temperature gradient at the border between the two parts of the boulder can be estimated as the temperature difference divided by the distance, $(\tau_\mathrm{w}-\tau_\mathrm{e})/l_\mathrm{ew}$, 
with $l_\mathrm{ew}$ being the typical distance between the eastern and western parts of the boulder,
i.e. roughly the distance between the centers of the two parts.
To estimate the last term in the right-hand side of Eqn. (\ref{heat_balance1}), $\tau$ can be substituted by its mean value, i.e. $\tau_\mathrm{w}$ and $\tau_\mathrm{e}$ for the western and the eastern parts of the boulder respectively.\footnote{When interpreted literally, the assumptions of a constant temperature gradient in the body of the boulder and of a constant temperature on the two parts of its surface might seem to contradict each other. Still, these assumptions should provide an acceptable estimate for the corresponding terms in Eqn. (\ref{heat_balance1}), and thus finally lead to a reasonable estimate for TYORP. 
It is possible to construct a more sophisticated model for the temperature distribution inside the boulder, but it will lead to a more complicated mathematics and a more obscure physics, while its accuracy will be anyway largely negated by the assumptions I am going to make below.}
Therefore, with these simplifications in place, Eqn. (\ref{heat_balance1}) for the two parts of the boulder assumes the following form:
\begin{eqnarray}
 \theta^2 v \tau_\mathrm{w}'&=& \frac{s_\mathrm{st}}{l_\mathrm{ew}}(\tau_\mathrm{e}-\tau_\mathrm{w})+s_\mathrm{sp}i_\mathrm{w}-s_\mathrm{sp}\tau_\mathrm{w}^4\,, \nonumber \\
 \theta^2 v \tau_\mathrm{e}'&=& \frac{s_\mathrm{st}}{l_\mathrm{ew}}(\tau_\mathrm{w}-\tau_\mathrm{e})+s_\mathrm{sp}i_\mathrm{e}-s_\mathrm{sp}\tau_\mathrm{e}^4\,.
 \label{heat_balance2}
\end{eqnarray}
Here $i_\mathrm{w}$ and $i_\mathrm{e}$ denote the dimensionless solar energy fluxes, defined as
\begin{eqnarray}
 i_\mathrm{w}&=& \frac{1}{\Phi s_\mathrm{sp}}\int\limits_{s_\mathrm{sp\ w}}I_i\mathrm{d}s_i\,, \nonumber \\
 i_\mathrm{e}&=& \frac{1}{\Phi s_\mathrm{sp}}\int\limits_{s_\mathrm{sp\ e}}I_i\mathrm{d}s_i\,.
\end{eqnarray}

I decompose $i_\mathrm{w}(\phi)$ and $i_\mathrm{e}(\phi)$ into a Fourier series,
and disregard all of the terms except for the zeroth and first order ones.
This simplification can alter the final result, but still can serve as an estimate.
Even with a modified illumination function, it is still a valid physical problem,
whose solution must still bear the basic properties of the tangential YORP.
Thus, for insolation I substitute
\begin{eqnarray}
 i_\mathrm{w}=C\cos{\phi}+S\sin{\phi}+\tau_0^4 \,, \nonumber \\
 i_\mathrm{e}=C\cos{\phi}-S\sin{\phi}+\tau_0^4 \,,
 \label{insolation}
\end{eqnarray}
where $C$, $S$, and $\tau_0$ are constants.
Here, $\tau_0$ has a physical meaning of the dimensionless temperature of the boulder, for which the emitted power equals the time-averaged absorbed power. Usually $\tau_0$ is close to the mean temperature of the boulder.
When writing the same $\tau_0$ in both equations, the same coefficients for cosine and opposite coefficients for sine,
I took into account that in the morning the eastern part of a symmetric boulder is illuminated in exactly the same manner
as the western part is in the evening, so that $i_\mathrm{w}(\phi)=i_\mathrm{e}(-\phi)$.

Next, I introduce the coefficients
\begin{equation}
 a=\frac{v}{s_\mathrm{sp}l}\,,\,\,\, b=\frac{s_\mathrm{st}l}{s_\mathrm{sp}l_\mathrm{ew}} \,,
 \label{ab}
\end{equation}
where $l$ is some typical boulder size.
Both coefficients $a$ and $b$ depend solely on the boulder shape, while the dependence on size only enters through $l$.

Finally, substituting Eqs. (\ref{insolation}) and (\ref{ab}) into Eqn. (\ref{heat_balance2}), I obtain
\begin{eqnarray}
 \theta^2 al \tau_\mathrm{w}'&=& \frac{b}{l} (\tau_\mathrm{e}-\tau_\mathrm{w})+C\cos{\phi}+S\sin{\phi}+\tau_0^4-\tau_\mathrm{w}^4\,,\nonumber \\
 \theta^2 al \tau_\mathrm{e}'&=& \frac{b}{l} (\tau_\mathrm{w}-\tau_\mathrm{e})+C\cos{\phi}-S\sin{\phi}+\tau_0^4-\tau_\mathrm{e}^4\,.
 \label{heat_balance}
\end{eqnarray}
Although this system is much simpler than the exact partial differential equation describing the heat conduction,
it still cannot be exactly solved analytically because of the nonlinearity $\tau^4$.
Therefore, I aim to construct its approximate analytic solution, which I do in the following subsection.

\subsection{Approximate solution of the heat conduction equations}
I am looking for the solution of Eqs. (\ref{heat_balance}) in the form of a series in terms of $C$ and $S$,
\begin{eqnarray}
 \tau_\mathrm{w}&=& \tau_\mathrm{w0}+\tau_\mathrm{w1}+\tau_\mathrm{w2}+...\,, \nonumber \\
 \tau_\mathrm{e}&=& \tau_\mathrm{e0}+\tau_\mathrm{e1}+\tau_\mathrm{e2}+...\,, 
 \label{decomposition}
\end{eqnarray}
where $\tau_\mathrm{w0}$ and $\tau_\mathrm{e0}$ are independent of $C$ and $S$,
$\tau_\mathrm{w1}$ and $\tau_\mathrm{e1}$ are proportional to the first powers of $C$ and $S$,
$\tau_\mathrm{w2}$ and $\tau_\mathrm{e2}$ are proportional to their second powers, and so on.
As I am looking for a relaxed periodic solution, all the terms have to be periodic in $\phi$ with a period of $2\pi$.
I am going to account for the contribution to TYORP of only the first three terms, and to disregard higher-order terms.
This is perfectly justified if $C,S\ll\tau_0^4$, but usually $C$ and $S$ are only slightly less than $\tau_0$ (compare with Table \ref{tab:geometry}).
It implies that the decomposition in the form of Eqn. (\ref{decomposition}) converges slowly, if at all.
Still, making $C$ and $S$ a factor of few smaller would make for a good convergence,
and all the further analysis would be justified.
Then I can assume that the extrapolation of the approximate formulas into the domain $C\sim S\sim \tau_0^4$ 
must give a reasonable order-of-magnitude estimate for TYORP.

Note, that Eqn. (\ref{decomposition}) is a Taylor series in terms of $C$ and $S$, 
in contrast to Eqn. (\ref{insolation}), which is a Fourier series in terms $\phi$.
I hold Fourier terms in Eqn. (\ref{insolation}) up to the first order and Taylor terms in Eqn. (\ref{decomposition}) up to the second order, in order to construct the minimal model for TYORP.
With merely zeroth-order terms in either Eqn. (\ref{insolation}) or Eqn. (\ref{decomposition}), TYORP would vanish.
Therefore, retaining first-order terms in both expansions is absolutely necessary. 
Retaining also the second-order Taylor term in Eqn. (\ref{decomposition}) is motivated by the fact that its contribution is of the same order as the contribution of the first-order Taylor term (as we will see later from Eqs. (\ref{2nd_2}) and (\ref{p_series})).
Taking more terms in either the Fourier decomposition Eqn. (\ref{insolation}) or the Taylor decomposition Eqn. (\ref{decomposition}) can make the solution more precise, but at the cost of increasing complexity of the problem and of the final expression.
Moreover, treating even an infinite number of terms in both the decompositions will not make the solution exact, because the initial Equation (\ref{heat_balance}) is already an approximation.
This can turn any attempt to go beyond the minimalistic model in Eqs. (\ref{insolation}) and (\ref{decomposition}) into an overkill.

Thus I substitute Eqn. (\ref{decomposition}) into Eqn. (\ref{heat_balance}),
and equate the terms of the same order. 
In the highest (zeroth) order I get
\begin{eqnarray}
 \theta^2 al \tau_\mathrm{w0}'&=& \frac{b}{l} (\tau_\mathrm{e0}-\tau_\mathrm{w0})+\tau_0^4-\tau_\mathrm{w0}^4\,, \nonumber \\
 \theta^2 al \tau_\mathrm{e0}'&=& \frac{b}{l} (\tau_\mathrm{w0}-\tau_\mathrm{e0})+\tau_0^4-\tau_\mathrm{e0}^4\,.
 \label{0th}
\end{eqnarray}
The periodic solution of this equation is $\tau_\mathrm{w0}=\tau_\mathrm{e0}=\tau_0$.

I substitute the obtained $\tau_\mathrm{w0}$ and $\tau_\mathrm{e0}$ back into Eqs. (\ref{heat_balance}) 
and write the terms of the first order, which are linear in terms of $C$ and $S$:
\begin{eqnarray}
 \theta^2 al \tau_\mathrm{w1}'&=& \frac{b}{l} (\tau_\mathrm{e1}-\tau_\mathrm{w1})+C\cos{\phi}+S\sin{\phi}-4\tau_0^3 \tau_\mathrm{w1} \,,\nonumber \\
 \theta^2 al \tau_\mathrm{e1}'&=& \frac{b}{l} (\tau_\mathrm{w1}-\tau_\mathrm{e1})+C\cos{\phi}-S\sin{\phi}-4\tau_0^3 \tau_\mathrm{e1} \,.
 \label{1st}
\end{eqnarray}
This is a system of linear differential equations with a sinusoidal inhomogeneity,
whose periodic solution can be found in the form
\begin{eqnarray}
\tau_\mathrm{w1}&=&C_\mathrm{w1}\cos{\phi}+S_\mathrm{w1}\sin{\phi} \,,\nonumber \\
\tau_\mathrm{e1}&=&C_\mathrm{e1}\cos{\phi}+S_\mathrm{e1}\sin{\phi} \,.
\label{sinusoidal_solution}
\end{eqnarray}
By substituting Eqn. (\ref{sinusoidal_solution}) into Eqn. (\ref{1st}) and equating the coefficients in front of the sines and cosines, I get
\begin{eqnarray}
 C_\mathrm{w1}&=&\frac{4\tau_0^3C}{16\tau_0^6+a^2l^2\theta^4}-\frac{al^3\theta^2S}{4(b+2l\tau_0^3)^2+a^2l^4\theta^4}\,,\nonumber \\
 S_\mathrm{w1}&=&\frac{al\theta^2C}{16\tau_0^6+a^2l^2\theta^4}+\frac{2l(b+2l\tau_0^3)S}{4(b+2l\tau_0^3)^2+a^2l^4\theta^4}\,,\nonumber \\
 C_\mathrm{e1}&=&\frac{4\tau_0^3C}{16\tau_0^6+a^2l^2\theta^4}+\frac{al^3\theta^2S}{4(b+2l\tau_0^3)^2+a^2l^4\theta^4}\,,\nonumber \\
 S_\mathrm{e1}&=&\frac{al\theta^2C}{16\tau_0^6+a^2l^2\theta^4}-\frac{2l(b+2l\tau_0^3)S}{4(b+2l\tau_0^3)^2+a^2l^4\theta^4}\,.
 \label{CS_coefficients}
\end{eqnarray}

Finally, I write down the second order terms of Eqs. (\ref{heat_balance}),
which are quadratic in terms of $C$ and $S$:
\begin{eqnarray}
 \theta^2 al \tau_\mathrm{w2}'&=& \frac{b}{l} (\tau_\mathrm{e2}-\tau_\mathrm{w2})-6\tau_0^2 \tau_\mathrm{w1}^2-4\tau_0^3 \tau_\mathrm{w2}\,, \nonumber \\
 \theta^2 al \tau_\mathrm{e2}'&=& \frac{b}{l} (\tau_\mathrm{w2}-\tau_\mathrm{e2})-6\tau_0^2 \tau_\mathrm{e1}^2-4\tau_0^3 \tau_\mathrm{e2}\,.
 \label{2nd}
\end{eqnarray}
I subtract Eqs. (\ref{2nd}) from each other, and average the result.
The left-hand side averages to 0, as $\tau_\mathrm{w2}$ and $\tau_\mathrm{e2}$ are periodic, and I am left with
\begin{equation}
 \langle\tau_\mathrm{w2}\rangle-\langle\tau_\mathrm{e2}\rangle
 =-\frac{3l\tau_0^2}{b+2l\tau_0^3} (\langle\tau_\mathrm{w1}^2\rangle-\langle\tau_\mathrm{e1}^2\rangle)\,.
\label{2nd_2}
\end{equation}
One does not need to find the exact expressions for $\tau_\mathrm{w2}$ and $\tau_\mathrm{e2}$:
as will be seen in the next subsection, Eqn. (\ref{2nd_2}) suffices to compute TYORP
in the second order in terms of $C$ and $S$.

\subsection{Computation of TYORP}
A heated surface emits light and experiences the recoil pressure
\begin{equation}
P=\frac{2}{3c}\epsilon\sigma T^4=\frac{(1-A)\Phi}{c}\frac{2}{3}\tau^4\,.
\label{pressure}
\end{equation}
Here $c$ is the speed of light, and the coefficient 2/3 corresponds to the light emission in accordance with Lambert's law.

To get the force experienced by the boulder in the eastward direction, I integrate this pressure over the boulder's surface, and take the $F_1$ component of the total force, assuming that the $x_1$ axis is directed from west to east:
\begin{equation}
F_1=-\int\limits_{S_\mathrm{sp}}P\,\mathrm{d}S_1\,.
\label{force}
\end{equation}

I average this force over time and nondimensionalize it by dividing it by $(1-A)\Phi S_\mathrm{proj}/c$,
where $S_\mathrm{proj}$ is the horizontal projected area of the boulder.
Thus I get the dimensionless TYORP pressure,
\begin{equation}
p=-\frac{2}{3}\frac{1}{s_\mathrm{proj}}\int\limits_{s_\mathrm{sp}}\langle\tau^4\rangle\,\mathrm{d}s_1\,.
\label{p_def}
\end{equation}
Now let us recall the assumption that the boulder is separated into two parts, 
with the dimensionless temperatures $\tau_\mathrm{w}$ and $\tau_\mathrm{e}$,
and the two parts are symmetric, with the same surface area of the western and eastern parts, 
$s_{\mathrm{sp}\,\mathrm{w}}=s_{\mathrm{sp}\,\mathrm{e}}$.
Assuming $\mathrm{d}s_1$ to always be negative at $s_{\mathrm{sp}\,\mathrm{w}}$ and always positive at $s_{\mathrm{sp}\,\mathrm{e}}$, I get
\begin{eqnarray}
p&=&-\frac{2}{3}\frac{1}{s_\mathrm{proj}}\left(
\langle\tau_\mathrm{w}^4\rangle\int\limits_{s_{\mathrm{sp}\,\mathrm{w}}}\mathrm{d}s_1
+\langle\tau_\mathrm{e}^4 \rangle\int\limits_{s_{\mathrm{sp}\,\mathrm{e}}}\mathrm{d}s_1\right)\nonumber \\
&=& \frac{2}{3}\frac{1}{s_\mathrm{proj}}\left(
\langle\tau_\mathrm{w}^4\rangle\int\limits_{s_{\mathrm{sp}\,\mathrm{w}}}\mathrm{d}|s_1|
-\langle\tau_\mathrm{e}^4 \rangle\int\limits_{s_{\mathrm{sp}\,\mathrm{e}}}\mathrm{d}|s_1|\right)\nonumber \\
&=&\frac{2}{3}\langle \tau_\mathrm{w}^4-\tau_\mathrm{e}^4\rangle\frac{1}{s_\mathrm{proj}}
\int\limits_{s_{\mathrm{sp}\,\mathrm{w}}}\mathrm{d}|s_1|\nonumber \\
&=&\frac{2}{3}\frac{s_\mathrm{ew}}{s_\mathrm{proj}}\langle \tau_\mathrm{w}^4-\tau_\mathrm{e}^4\rangle\,.
\label{p_transformation}
\end{eqnarray}
Here $s_\mathrm{ew}=\int_{s_{\mathrm{sp}\,\mathrm{w}}}\mathrm{d}|s_1|$ denotes the area of the boulder projected onto the vertical meridianal plane. It is convenient to introduce a new coefficient,
\begin{equation}
n=\frac{2s_\mathrm{ew}}{s_\mathrm{proj}}\,.
\label{n}
\end{equation}
This coefficient $n$ characterizes steepness of the surface. If $s_\mathrm{proj}$ is understood as the horizontal projected area of the boulder, then $n=\langle\tan{\alpha_\mathrm{ew}}\rangle$ is the mean tangent of the slope of the surface in the east-west direction.
It is zero for a flat surface, unity for a surface with 45$^\circ$ slopes, and bigger for even steeper slopes. 
Now, substituting Eqn. (\ref{n}) into Eqn. (\ref{p_transformation}) 
and using the decomposition of $\tau$ from Eqn. (\ref{decomposition}), results into
\begin{eqnarray}
p&=&\frac{1}{3}n\langle \tau_\mathrm{w}^4-\tau_\mathrm{e}^4 \rangle \nonumber \\
&=&\frac{1}{3}n\langle \tau_0^4+4\tau_0^3\tau_\mathrm{w1}+4\tau_0^3\tau_\mathrm{w2}+6\tau_0^2\tau_\mathrm{w1}^2+... \nonumber \\
&&-\tau_0^4-4\tau_0^3\tau_\mathrm{e1}-4\tau_0^3\tau_\mathrm{e2}-6\tau_0^2\tau_\mathrm{e1}^2-... \rangle\,.
\label{p_series}
\end{eqnarray}
As $\tau_\mathrm{w1}$ and $\tau_\mathrm{e1}$ are sinusoidal, their means are 0.
The means of $\tau_\mathrm{w2}$ and $\tau_\mathrm{e2}$ are obtained from Eqn. (\ref{2nd_2}). 
Thus Eqn. (\ref{p_series}) results in
\begin{eqnarray}
p=\frac{2nb\tau_0^2}{b+2l\tau_0^3} (\langle\tau_\mathrm{w1}^2\rangle-\langle\tau_\mathrm{e1}^2\rangle)\,.
\label{p_tau1}
\end{eqnarray}
Using Eqs. (\ref{sinusoidal_solution}) for $\tau_\mathrm{w1}$ and $\tau_\mathrm{e1}$ leads to
\begin{eqnarray}
\langle\tau_\mathrm{w1}^2\rangle-\langle\tau_\mathrm{e1}^2\rangle=\frac{1}{2}(C_\mathrm{w1}^2+S_\mathrm{w1}^2-C_\mathrm{e1}^2-S_\mathrm{e1}^2)\,.
\end{eqnarray}
Finally, substituting this expression into Eqn. (\ref{p_tau1}) and using Eqn. (\ref{CS_coefficients}), I obtain
\begin{eqnarray}
p=\frac{8nab^2\tau_0^2l^2\theta^2CS}{(b+2l\tau_0^3)(16\tau_0^6+a^2l^2\theta^4)(4(b+2l\tau_0^3)^2+a^2l^4\theta^4)}\,.
\label{p_general}
\end{eqnarray}

This is the analytic expression of TYORP that I was looking for. It allows one to approximately estimate TYORP using the shape and thermal properties of the boulder.
In the following section, I apply it to several different boulder shapes.

\section{Application of analytic TYORP expression to boulders of different shapes\label{sec-shapes}}
\subsection{One-dimensional wall}
\begin{figure}
\centering
\begin{tikzpicture}
\draw [color=gray!20,fill=gray!20] (-4,1)--(-1,1)--(-1,-0.5)--(-4,-0.5)--cycle;
\draw [color=gray!20,fill=gray!20] (4,1)--(0,1)--(0,-0.5)--(4,-0.5)--cycle;
\draw [color=gray!20,fill=gray!20] (-4,-2)--(-3,-2.5)--(-1,-1.5)--(1,-2.5)--(3,-1.5)--(4,-2.)--(4,-3.5)--(-4,-3.5)--cycle;
\draw [color=gray!50,fill=gray!50] (-3,1)--(-2,1)--(-2,3)--(-3,3)--cycle;

\draw[thick] (-4,1) -- (-1,1);
\draw[thick] (0,1) -- (4,1);
\draw[thick] (-4,-2.) -- (-3,-2.5);
\draw[thick] (-3,-2.5) -- (-1,-1.5);
\draw[thick] (-1,-1.5) -- (1,-2.5);
\draw[thick] (1,-2.5) -- (3,-1.5);
\draw[thick] (3,-1.5) -- (4,-2.);

\draw[gray] (-1,-1.5)--(3,-1.5);
\draw[gray] (1,-2.5)--(2.5,-2.5);

\draw[thick] (-2,1)--(-2,3);
\draw[thick] (-2,3)--(-3,3);
\draw[thick] (-3,1)--(-3,3);

\fill[gray!50] (2,1) circle [radius=1.];
\draw (2,1) circle [radius=1.];
\draw[gray] (2,1)--(1.295,0.295);

\draw[dashed] (-2.5,1)--(-2.5,3);
\draw[dashed] (2,0)--(2,2);
\draw[dashed] (-3,-2.5)--(-3,-3);
\draw[dashed] (-1,-1.5)--(-1,-3);
\draw[dashed] (1,-2.5)--(1,-3);
\draw[dashed] (3,-1.5)--(3,-3);
\draw[dashed] (-4,-3)--(4,-3);
\draw[gray] (4,-2)--(4,-3);

\node[] at (-2.5,0.7) {$d$};
\node[] at (-3.2,2) {$s$};
\node[] at (1.5,0.8) {$r$};
\node[] at (1,-1.3) {$\lambda$};
\node[] at (1.7,-2.35) {$\alpha$};
\node[] at (3.8,-2.5) {$\frac{\lambda}{2\pi}$};
\node[] at (-2,-2.7) {$\frac{\lambda}{2}$};

\node[] at (-3.7,2.7) {(a)};
\node[] at (0.3,2.7) {(b)};
\node[] at (-3.7,-1.3) {(c)};
\end{tikzpicture}
\caption{Geometries, used for modeling: (a) wall; (b) ball; (c) wave.}
\label{fig:geometries}%
\end{figure}
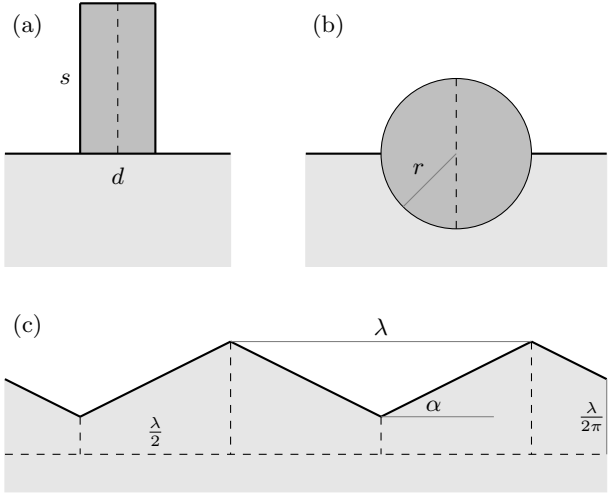

Let us start testing this analytic expression with the simplest and historically first model of TYORP:
one-dimensional heat conductivity in a wall by \cite{golubov12}.
The high long wall stands on regolith, going from the north to the south, and is illuminated by the sun alternatively from the east or the west (panel (a) in Figure \ref{fig:geometries}).
Given that the dimensionless thickness of the wall is $d$ and considering its patch of dimensionless surface area $s$, I separate the wall into the eastern and the western slabs with the thicknesses $\frac{d}{2}$. Then the volume of each slab is $v=\frac{sd}{2}$ and the typical distance between the slabs is $l_\mathrm{ew}=\frac{d}{2}$. Thus I go on, filling in the geometric properties in the first row of Table \ref{tab:geometry}, until I get $a$, $b$ and $n$. (Note though, that for $s_\mathrm{proj}$ I take not a horizontal projected surface area, but a vertical one, which is the only meaningful definition for a very high vertical wall. It also renders $n$ different from its conventional meaning of $\langle\tan{\alpha_\mathrm{ew}}\rangle$.)

To compute the last three columns in Table \ref{tab:geometry}, I need the insolation function, for which I take
\begin{equation}
 i_w(\phi)=2\mathrm{sin}\phi,\, 0<\phi<\frac{\pi}{2}.
 \label{i_wall}
\end{equation}
The coefficient 2 is due to the assumed mirror reflection of light by the regolith.
Decomposing $i_w(\phi)$ into the Fourier series, I finish filling in the first row of Table \ref{tab:geometry}.

\begin{table*}
\centering
\caption{Geometric properties of different boulder shapes}
\begin{tabular}{lccccccccccccc}
Shape & $l$ & $v$ & $s_\mathrm{st}$ & $s_\mathrm{sp}$ & $s_\mathrm{proj}$ & $s_\mathrm{ew}$ & $l_\mathrm{ew}$ & $a=\frac{v}{s_\mathrm{sp}l}$ & $b=\frac{s_\mathrm{st}l}{s_\mathrm{sp}l_\mathrm{ew}}$ & $n=\frac{2s_\mathrm{ew}}{s_\mathrm{proj}}$ & $\tau_0$ & $C$ & $S$ \\ 
\hline
Wall & $d$ & $\frac{sd}{2}$ & $s$ & $s$ & $s$ & $s$ & $\frac{d}{2}$ & $\frac{1}{2}$ & 2 & 2 & $\frac{1}{\sqrt[4]{\pi}}$ & $\frac{1}{\pi}$ & $\frac{1}{2}$ \\
Sphere (mirror) & $r$ & $\frac{2}{3}\pi r^3$ & $\pi r^2$ & $\pi r^2$ & $\pi r^2$ & $\frac{1}{2}\pi r^2$ & $r$ & $\frac{2}{3}$ & 1 & 1 & $\frac{1}{\sqrt{2}}$ & $\frac{1}{\pi}$ & $\frac{1}{4}$\\
Sphere (absorbing) & $r$ & $\frac{2}{3}\pi r^3$ & $\pi r^2$ & $\pi r^2$ & $\pi r^2$ & $\frac{1}{2}\pi r^2$ & $r$ & $\frac{2}{3}$ & 1 & 1 & $\sqrt[4]{\frac{\pi+2}{8\pi}}$ & $\frac{\pi+4}{8\pi}$ & $\frac{\pi-1}{4\pi}$\\
Wave & $\lambda$ & $\frac{\lambda s}{2\pi}$ & $\frac{s}{\pi}$ & $\frac{s}{2}$ & $s$ & $\frac{s\alpha}{2}$ & $\frac{\lambda}{2}$ & $\frac{1}{2\pi}$ & $\frac{4}{\pi}$ & $\alpha$ & $\frac{1}{\sqrt[4]{\pi}}$ & $\frac{1}{2}$ & $\frac{\alpha}{2}$ \\ 
\end{tabular}
\label{tab:geometry}
\end{table*}

\begin{figure*}
  \centering    
  \includegraphics[width=0.45\textwidth]{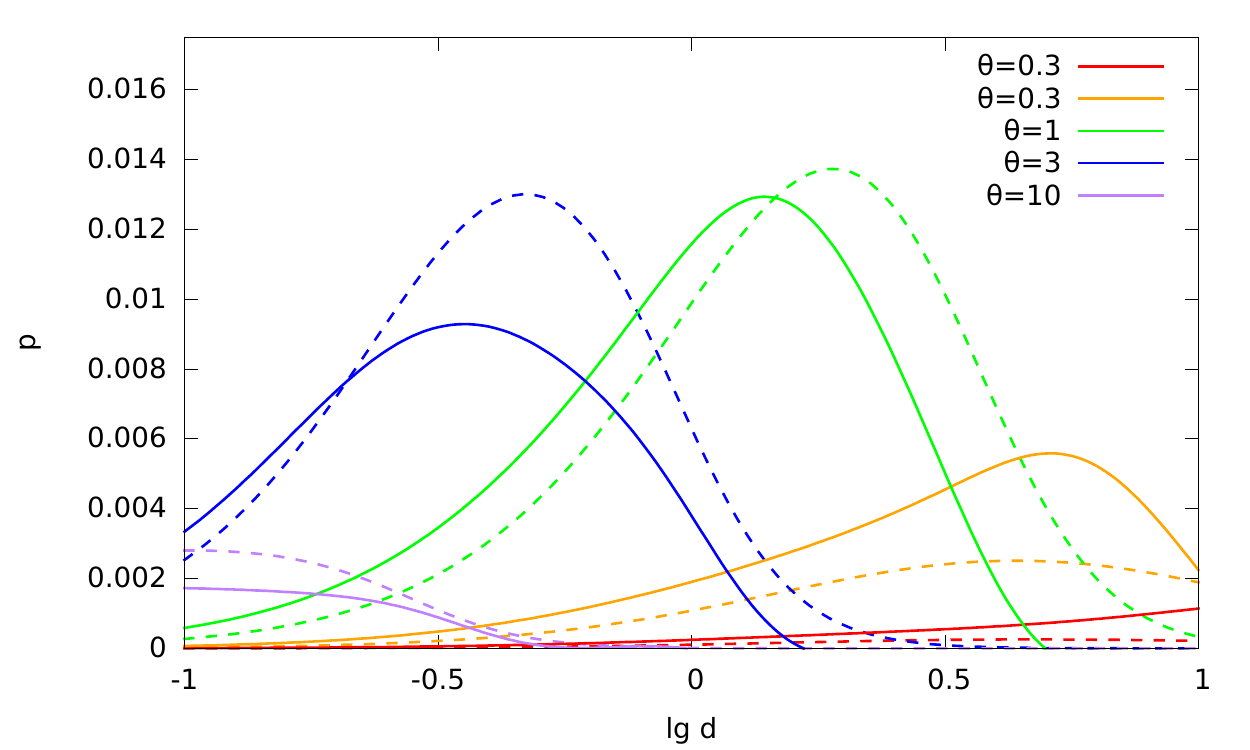}
  \includegraphics[width=0.45\textwidth]{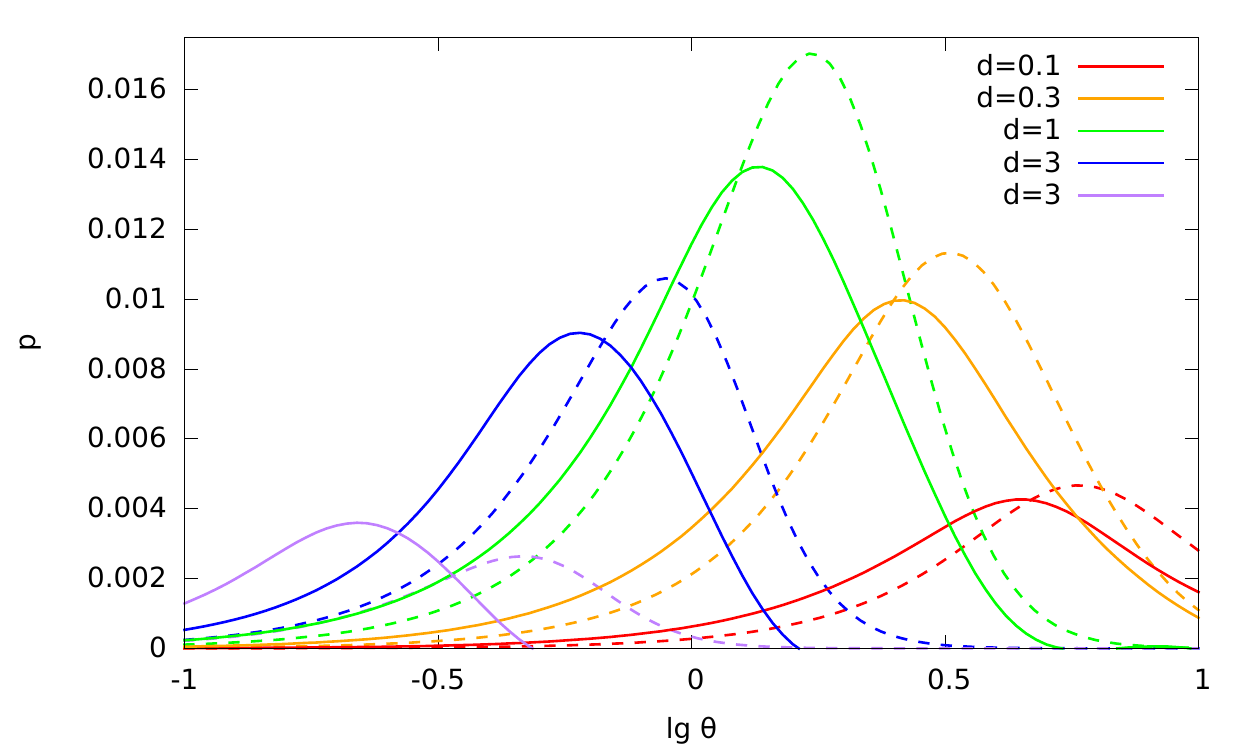}
  \caption{The dimensionless pressure for a wall.
  Solid lines represent the results of numeric simulations \citep{golubov11,golubov12},
  dashed lines are computed with the analytic formula Eqn. \ref{p_general}.
  \textit{Left:} Dimensionless pressure as a function of the dimensionless thickness of the wall $d$ for different values of the thermal parameter $\theta$.
  \textit{Right:} Dimensionless pressure as a function of $\theta$ for different values of $d$.}
  \label{fig:wall_plots}
\end{figure*}

\begin{figure*}
  \centering    
  \includegraphics[width=0.45\textwidth]{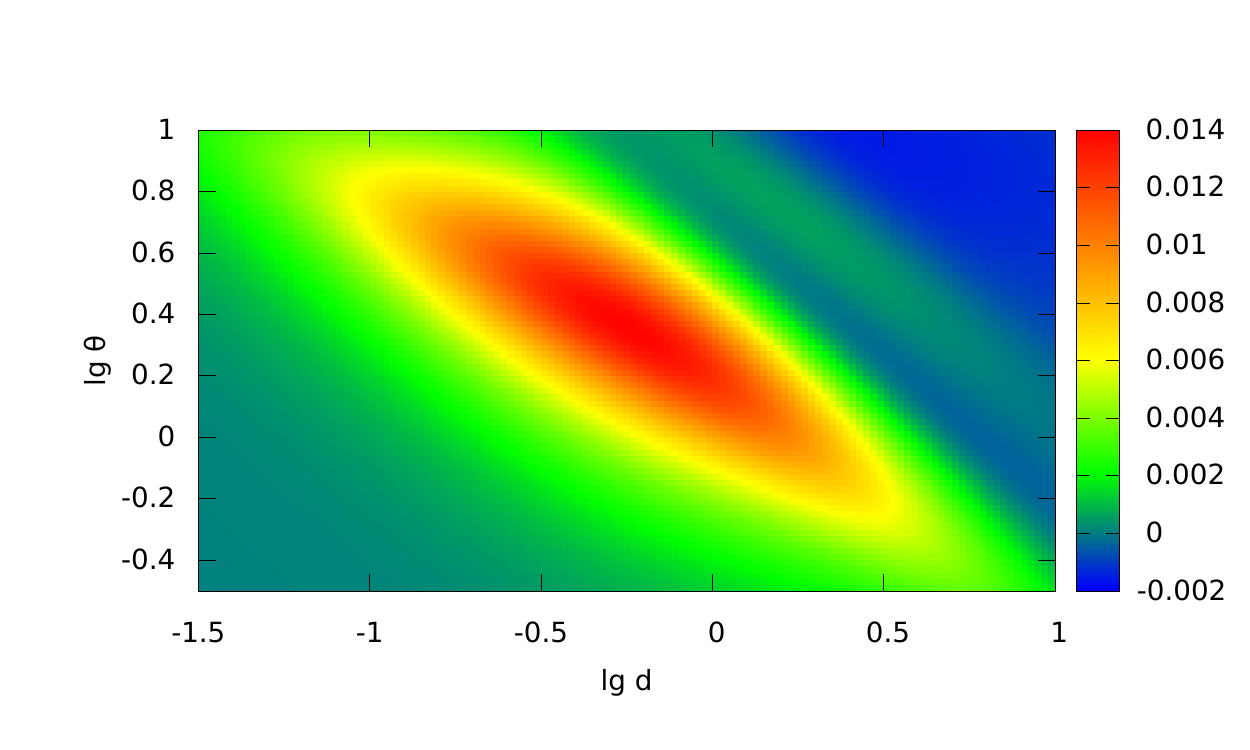}
  \includegraphics[width=0.45\textwidth]{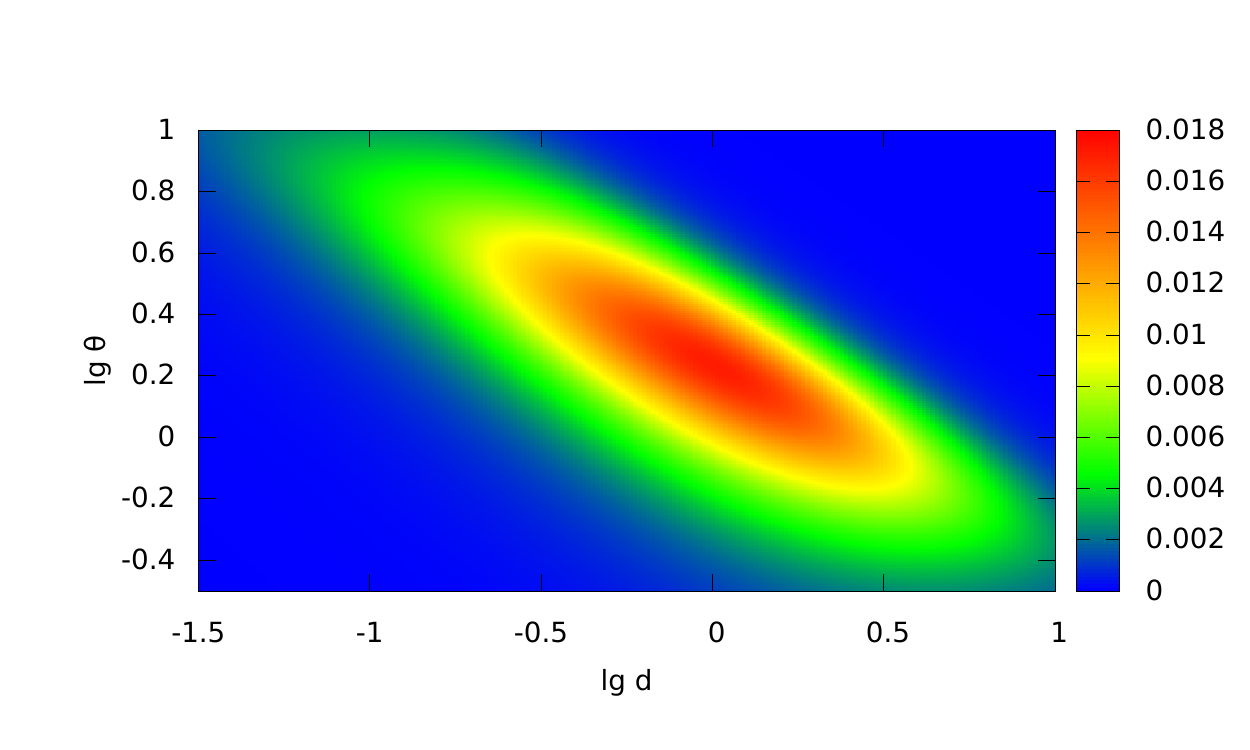}
  \caption{The dimensionless pressure for a wall color-coded as a function of the wall thickness $d$ and the thermal parameter $\theta$.
  The left-hand panel was obtained from numeric simulations \citep{golubov12},
  while the right-hand panel was computed with the analytic formula Eqn. \ref{p_general}.}
  \label{fig:wall_maps}
\end{figure*}

Now I substitute the obtained coefficients $a$, $b$, $n$, $\tau_0$, $C$ and $S$ into Eqn. \ref{p_general}, and in Figures \ref{fig:wall_plots} and \ref{fig:wall_maps} compare the resulting analytic expression for TYORP pressure $p$ with the numeric simulations by \cite{golubov11} and \cite{golubov12}.
The agreement is almost too good, given the approximations made while constructing the theory.
For the values of $l$ and $\theta$ that correspond to big $p$, the value of $p$ is predicted with an accuracy of a few tens per cent (Figure \ref{fig:wall_plots}), although far from the maximal $p$ the accuracy is worse.
The area where $p$ is big, is very similar in the analytic theory and in the simulations (Figure \ref{fig:wall_plots}).
The qualitative behavior of $p$ as a function of $l$ and $\theta$ is also reproduced correctly.
This is as good as one could expect from a simple estimate done by such an approximate theory.

A few warnings are still to be made.
In Figure \ref{fig:wall_plots}, I disregard negative simulated values of $p$, observed by \cite{golubov12} for big $l$ and $\theta$.
\cite{golubov12} suggested that these negative values could be numeric artifacts,
and this was later confirmed by \cite{sevecek15}, who only obtained positive $p$ in all their simulations.
Furthermore, \cite{sevecek15} proved that mirror reflection of light by the regolith is a bad approximation (worse, in fact, than disregarding reflected light completely).
Still, this approximation enters both the theory and the simulations, 
so they should be wrong in the same way, and thus agree with each other, which they do.

\subsection{Spherical boulder}
The second model I want to study is a particular case from \cite{golubov14}:
a spherical boulder of radius $r$ lying half buried in the regolith on the equator of an asteroid (panel (b) in Figure \ref{fig:geometries}).
I separate the boulder into the eastern and the western hemispheres,
and list all the necessary geometric properties in Table \ref{tab:geometry}.
For the typical distance between the hemispheres $l_\mathrm{ew}$ I take $r$, which is half the distance between their most remote points. Measuring the distance between their centers of mass instead would result into $\frac{3}{4}r$, and an argument could be also made for using this value instead, although the difference should not matter much given the crudeness of all the previously made assumptions.
For $s_\mathrm{proj}$ I take the horizontal projection of the boulder, $\pi r^2$.
This is different from \cite{golubov14}, who considered a regular array of boulders and used for $s_\mathrm{proj}$ the surface area per boulder. Therefore, I re-normalize their results. Note that the choice of $s_\mathrm{proj}$ does not influence the observable physical values: $s_\mathrm{proj}$ enters the definition of $p$ inversely, but also enters the transformation factor from $p$ to $F$ directly, and thus cancels out.

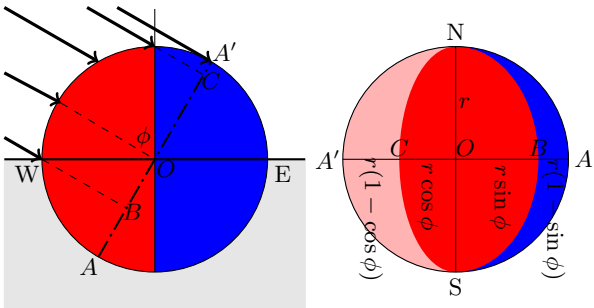
\begin{figure}
\centering
\begin{tikzpicture}
\draw [color=gray!20,fill=gray!20] (-4,0)--(0,0)--(0,-2)--(-4,-2)--cycle;
\fill[red] (-2,-1.5) arc(270:90:1.5) --cycle;
\fill[blue] (-2,-1.5) arc(-90:90:1.5) --cycle;
\draw (-2,0) circle [radius=1.5];
\draw[thick] (-4,0)--(0,0);
\draw (-2,-1.5)--(-2,2);
\draw[thick, dash pattern={on 7pt off 2pt on 1pt off 2pt}] (-2.75,-1.3)--(-1.25,1.3);

\draw[dashed] (-3.5,0)--(-2.374,-0.65);
\draw[->,very thick] (-4, 0.29)--(-3.5,0);
\draw[dashed] (-3.25,0.75)--(-2,0);
\draw[->,very thick] (-4,1.15)--(-3.25,0.75);
\draw[->,very thick] (-4, 2.02)--(-2.75,1.3);
\draw[dashed] (-2,1.5)--(-1.35,1.125);
\draw[->,very thick] (-2.9,2.02)--(-2,1.5);
\draw[->,very thick] (-2.5,2.02)--(-1.25,1.3);

\node[] at (-2.15,0.3) {$\phi$};
\node[] at (-3.7,-0.2) {W};
\node[] at (-0.3,-0.2) {E};
\node[] at (-1.07,1.43) {$A'$};
\node[] at (-2.88,-1.43) {$A$};
\node[] at (-2.3,-0.7) {$B$};
\node[] at (-1.85,-0.13) {$O$};
\node[] at (-1.25,1.05) {$C$};

\fill[blue] (2,0) circle [radius=1.5];
\fill[red] (2,0) ellipse (1.1 and 1.5);
\fill[red!30] (2,1.5) arc(90:270:1.5) --cycle;
\fill[red] (2,0) ellipse (0.75 and 1.5);
\draw (2,0) circle [radius=1.5];
\draw (0.5, 0)--(3.5,0);
\draw (2,-1.5)--(2,1.5);
\node[rotate=270] at (2.6,-0.5) {$r\sin\phi$};
\node[rotate=270] at (1.63,-0.5) {$r\cos\phi$};
\node[rotate=270] at (3.3,-0.8) {$r(1-\sin\phi)$};
\node[rotate=270] at (0.875,-0.8) {$r(1-\cos\phi)$};
\node[] at (2.12,0.75) {$r$};
\node[] at (2,1.7) {N};
\node[] at (2,-1.7) {S};
\node[] at (3.7,0) {$A$};
\node[] at (0.3,0) {$A'$};
\node[] at (3.1,0.15) {$B$};
\node[] at (2.13,0.15) {$O$};
\node[] at (1.25,0.15) {$C$};

\end{tikzpicture}
\caption{Illustration of illumination of a spherical boulder. 
\textit{Left:} Solar rays falling onto the boulder. The western side of the boulder is shown in red, the eastern side is shown in blue. 
\textit{Right:} Illuminated areas in the eastern and western halves of the boulder, projected onto the plane $AA'$, normal to the incident light. The letters N, E, S, W denote cardinal directions. The area, from which solar rays come directly to the western side of the boulder is shown in red, directly to the eastern side of the boulder -- in blue, come to the western side of the boulder after mirror reflection from the regolith -- in pink. To get $i_w$ for light-absorbing regolith, we must divide the red-colored area by $s_\mathrm{sp}=\pi r^2$. For light-reflecting regolith, the pink-colored part of the circle should be added.}
\label{fig:sphere-illumination}%
\end{figure}

For the insolation I try two possibilities.
First, assuming mirror reflection of light by the regolith, I get (see Figure \ref{fig:sphere-illumination} for explanation)
\begin{equation}
 i_w(\phi)=\frac{1}{2}(1+\mathrm{sin}\phi),\, -\frac{\pi}{2}<\phi<\frac{\pi}{2}.
 \label{i_ball1}
\end{equation}
Second, assuming full absorption of light by the regolith, and thus accounting for only the direct solar irradiation, I get
\begin{eqnarray}
 i_w(\phi)=\frac{1}{2}(1+\mathrm{sin}\phi),\,\,\mathrm{if}\,\, -\frac{\pi}{2}<\phi<0, \nonumber\\
 i_w(\phi)=\frac{1}{2}(\mathrm{cos}\phi+\mathrm{sin}\phi),\,\,\mathrm{if}\,\, 0<\phi<\frac{\pi}{2}.
 \label{i_ball2}
\end{eqnarray} 
I treat the two cases separately in the second and third rows of Table \ref{tab:geometry}.

\begin{figure*}
  \centering    
  \includegraphics[width=0.45\textwidth]{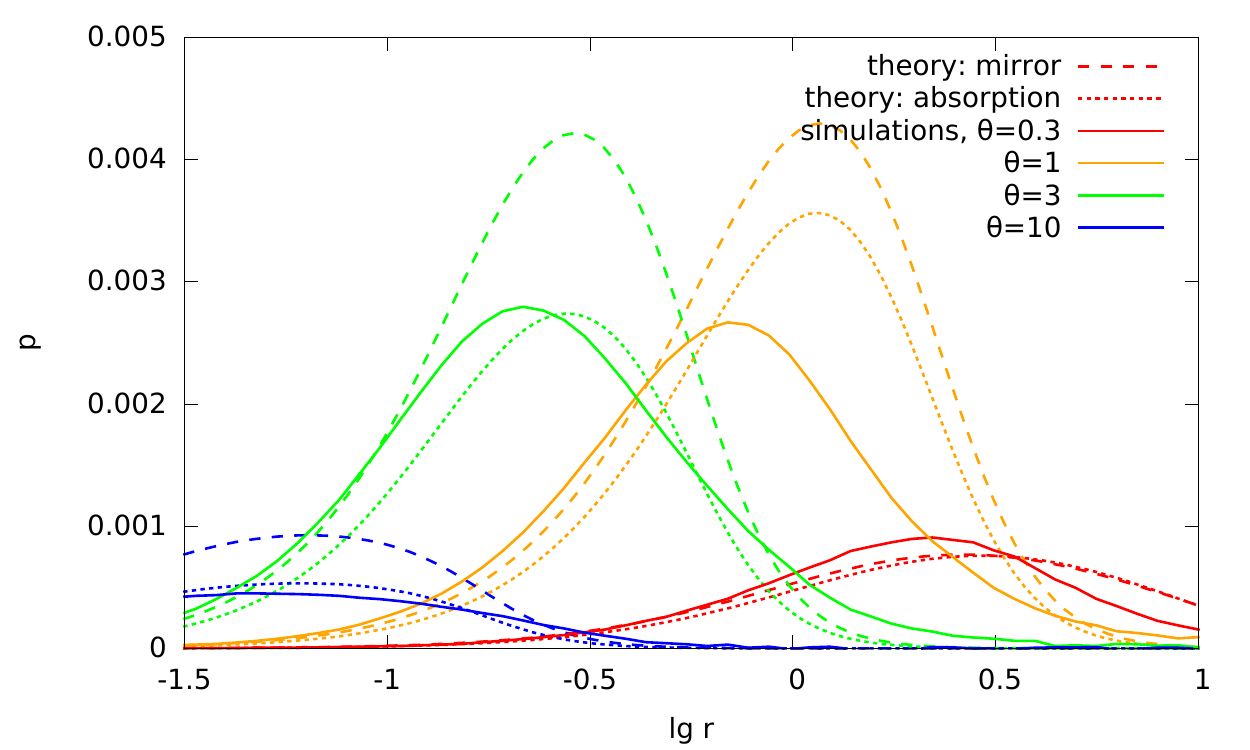}
  \includegraphics[width=0.45\textwidth]{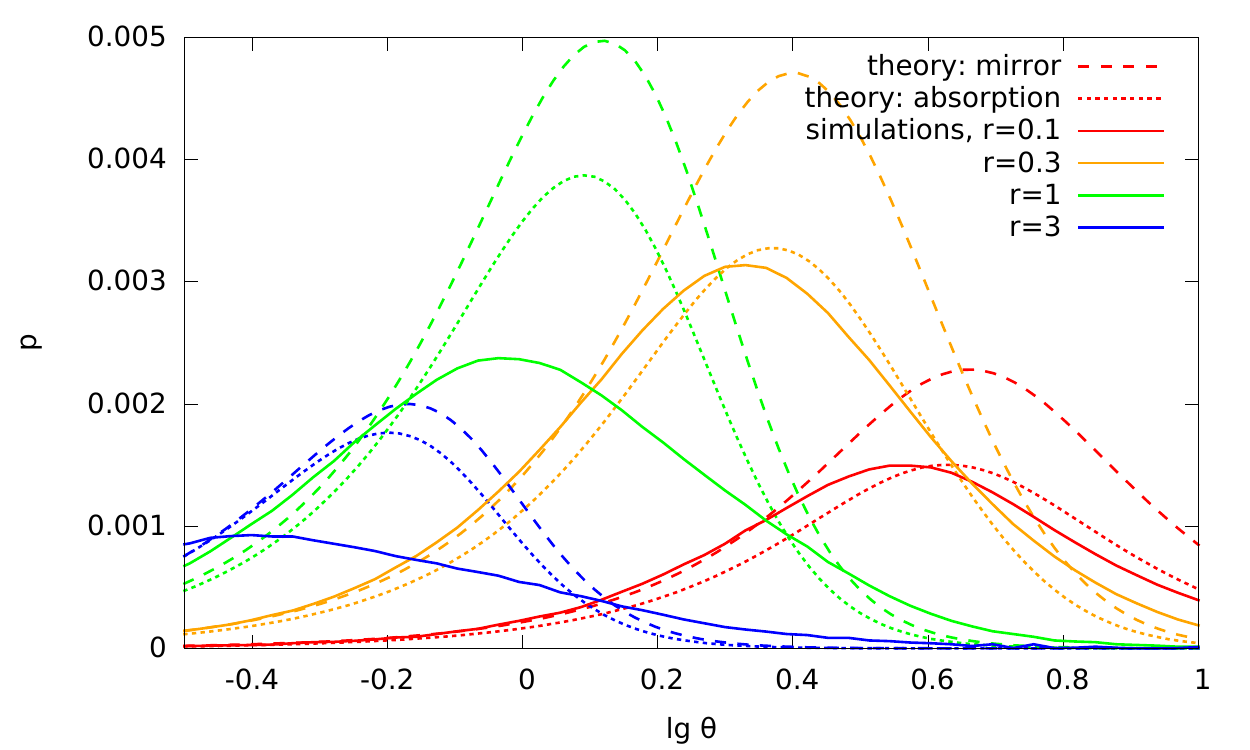}
  \caption{The dimensionless pressure for a spherical boulder.
  Solid lines represent the results of numeric simulations \citep{golubov14},
  dashed lines are computed with the analytic formula Eqn. \ref{p_general}.
  \textit{Left:} Dimensionless pressure as a function of the boulder radius $r$ for different values of the thermal parameter $\theta$.
  \textit{Right:} Dimensionless pressure as a function of $\theta$ for different values of $r$.}
  \label{fig:sphere_plots}
\end{figure*}

In Figure \ref{fig:sphere_plots}, I compare the numeric simulations of \citep{golubov14} (solid lines)
with the two analytic models.
Qualitatively both of them work well, but the model including only the direct light has a better quantitative agreement with the simulations than the model with mirror reflection (similarly to findings of \cite{sevecek15} for a wall).
In Figure \ref{fig:wall_maps}, I compare the better analytical model with the numeric simulation.
One must also keep in mind that the simulations of \citep{golubov14} were conducted for an array of boulders partially shadowing each other.
Subtracting the shadowing should increase TYORP by a few tens per cent, as seen in the middle right panel of Figure 4 in \citep{golubov14}.
This could somewhat improve the agreement between the presented analytical model and the simulations.

\begin{figure*}
  \centering    
  \includegraphics[width=0.45\textwidth]{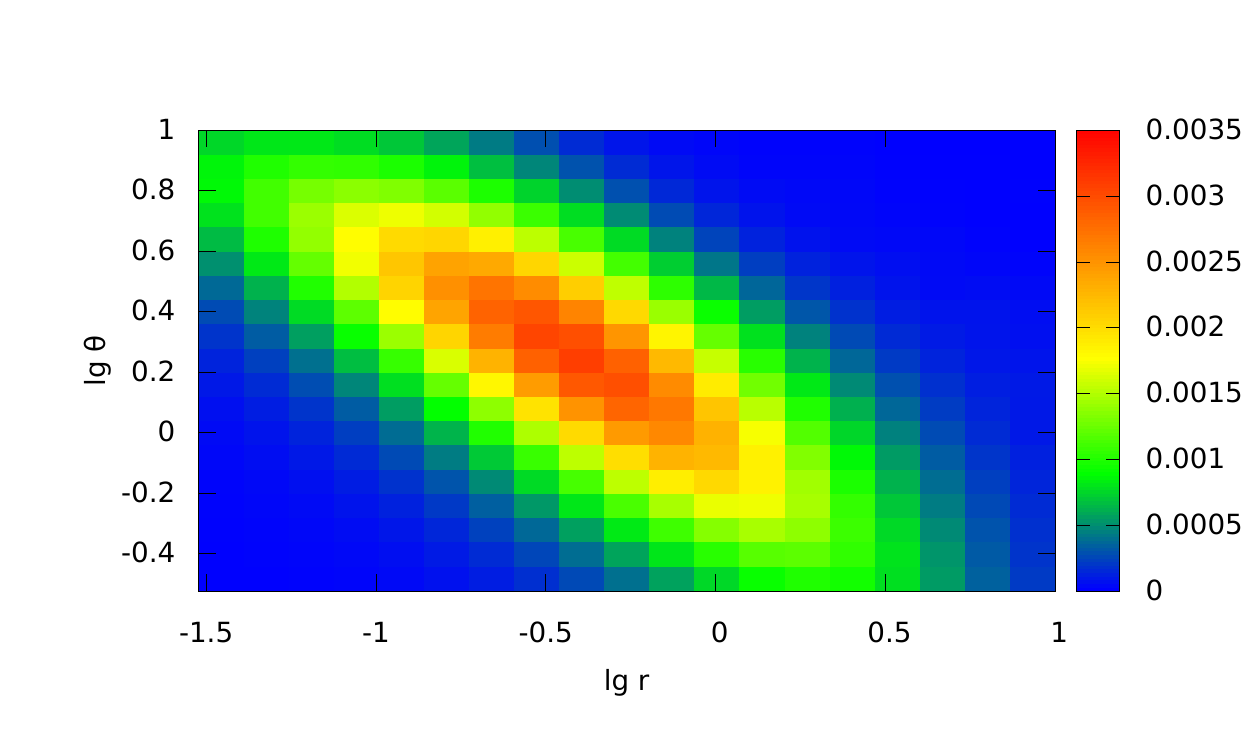}
  \includegraphics[width=0.45\textwidth]{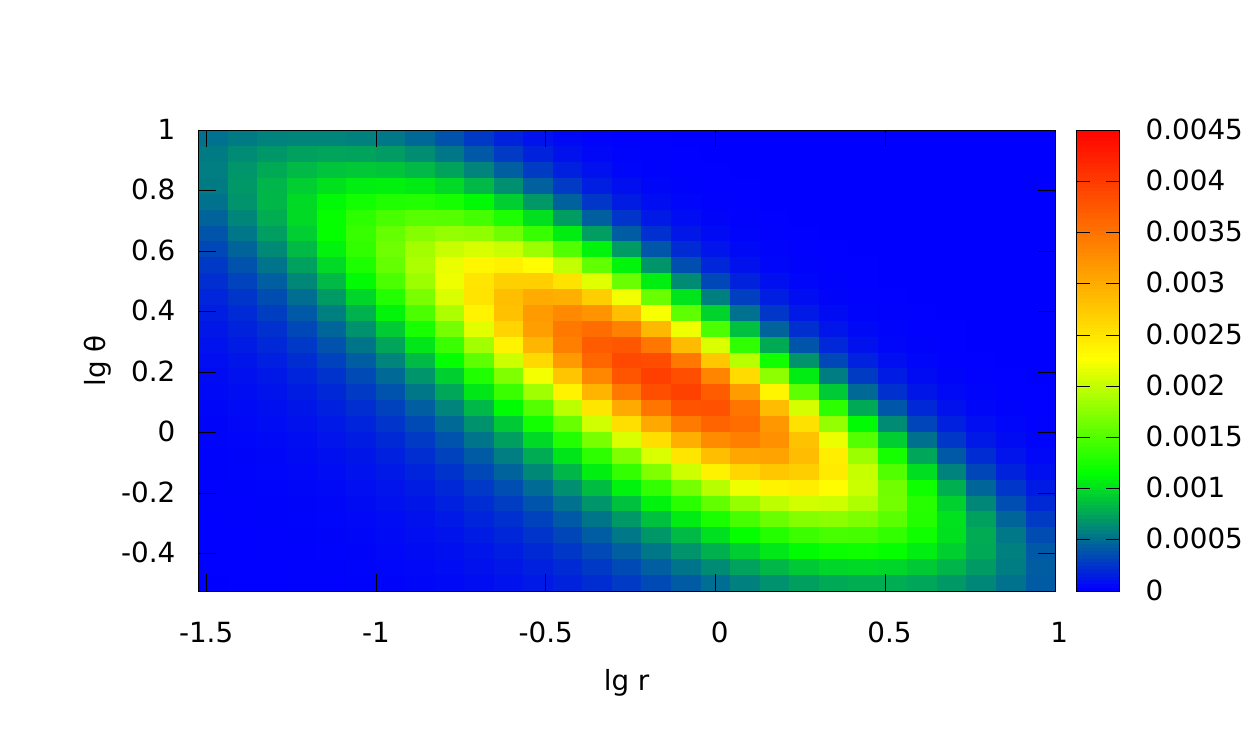}
  \caption{The dimensionless pressure for a spherical boulder color-coded as a function of the boulder radius $r$ and the thermal parameter $\theta$.
  The left-hand panel was obtained from numeric simulations \citep{golubov14},
  while the right-hand panel was computed with the analytic formula Eqn. \ref{p_general}, accounting only for direct solar light (third row in Table \ref{tab:geometry}).}
  \label{fig:sphere_maps}
\end{figure*}

\subsection{Wave in the regolith}

Now, having tested the model for already studied cases, let us move to an as-yet unknown terrain.
Consider a regular array of hills and dales, with flat slopes angled at $\alpha$, with a dimensionless wavelength $\lambda$,
all positioned on the equator in the north-south direction (panel (c) in Figure \ref{fig:geometries}).
The model is intended as a proxy for small bumps and pits on the regolith, whose contribution to TYORP has never yet been considered.
What I am going to do here is but a crude estimate, yet to be tested by thorough numeric simulations.
Unlike for a boulder with a well-specified volume, now the volume in which heat conduction occurs has no well-defined borders, and estimating geometric properties for Table \ref{tab:geometry} gets more complicated. Obviously, considering an infinitely deep volume below the surface is unsatisfactory, as an infinitely large volume will have infinite heat capacity, zero temperature oscillations, and thus zero TYORP.

Differently directed slopes cause uneven heating of the surface, and thus temperature variations in the east-west direction.
The principal Fourier harmonics of this horizontal temperature oscillation has the wavelength $\lambda$, and the corresponding wave vector $k=\frac{2\pi}{\lambda}$. The temperature oscillation will dampen with depth, and one can estimate the penetration depth of these temperature oscillations to be of the order of $\frac{1}{k}=\frac{\lambda}{2\pi}$.
Thus I consider heat conduction in a slab of depth $\frac{\lambda}{2\pi}$ and surface area $s$.
Although for $\lambda$ or $\theta$ very different from unity the typical depth can be different and determined by other physical processes,
and thus the model can produce a bigger error than in the previous two cases,
even now it must at least give a reasonable order-of-magnitude estimate of TYORP for $\lambda\sim\theta\sim 1$.
The slab under consideration borders two other slabs, to the east and to the west.
These two slabs have the same temperature, and one can account for having two of them by just doubling their area of contact, $s_\mathrm{st}$.

Having said all this, I fill in the last row in Table \ref{tab:geometry}.
I assume that $\alpha$ is small, thus substituting $\mathrm{cos}\alpha\approx 1$, $\mathrm{tan}\alpha\approx\mathrm{sin}\alpha\approx\alpha$.
For the illumination function I take
\begin{equation}
 i_w(\phi)=\mathrm{cos}(\phi-\alpha),\, -\frac{\pi}{2}+\alpha<\phi<\frac{\pi}{2}-\alpha.
 \label{i_wall}
\end{equation}
When Fourier-decomposing $i_w(\phi)$, I also keep only the principal terms in terms of $\alpha$.

\begin{figure*}
  \centering    
  \includegraphics[width=0.45\textwidth]{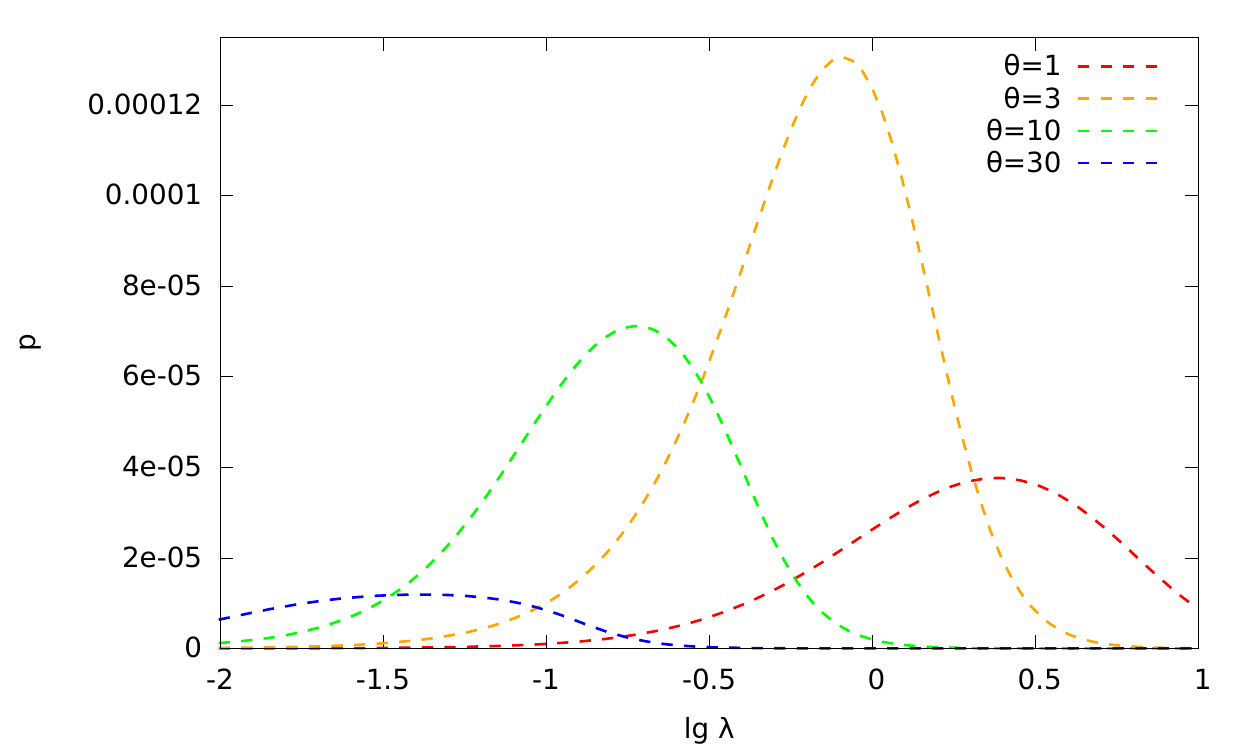}
  \includegraphics[width=0.45\textwidth]{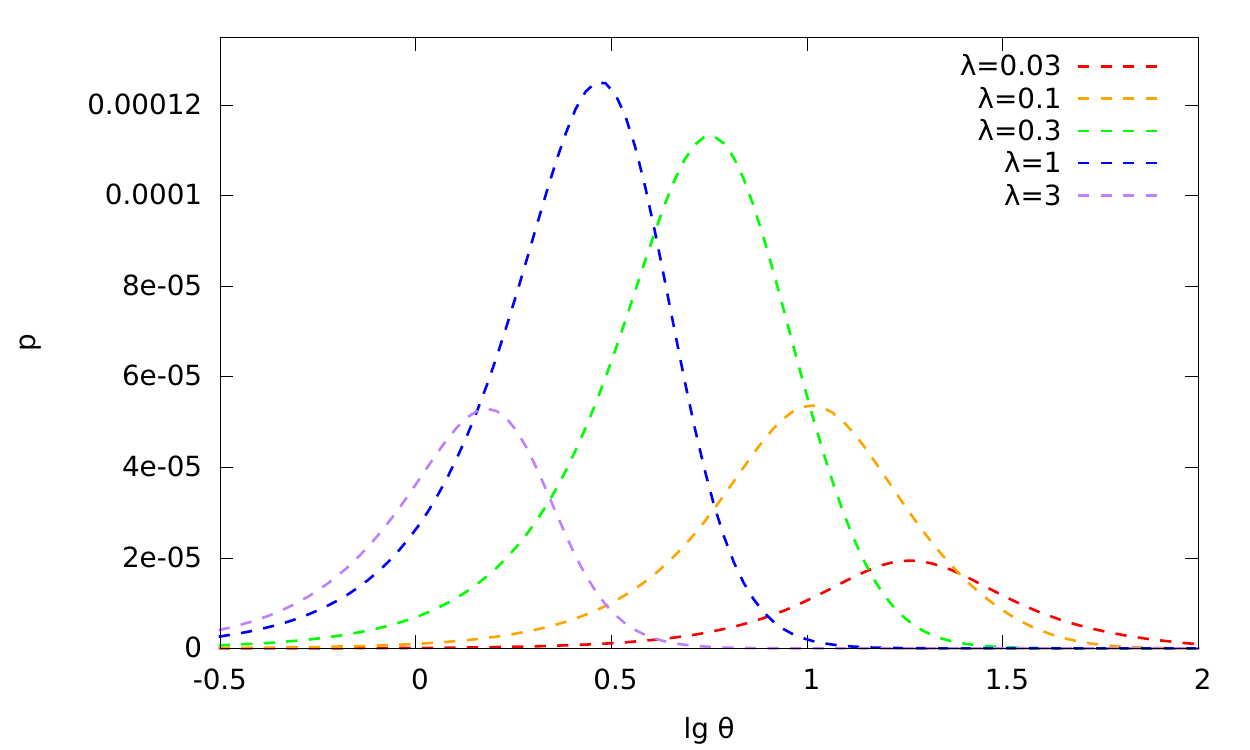}
  \caption{The dimensionless pressure for a wave with the slope $\alpha=0.1$, computed with the analytic formula Eqn. \ref{p_general}.
  \textit{Left:} Dimensionless pressure as a function of the dimensionless wavelength $\lambda$ for different values of the thermal parameter $\theta$.
  \textit{Right:} Dimensionless pressure as a function of $\theta$ for different values of $\lambda$.}
  \label{fig:wave_plots}
\end{figure*}

Substituting the last row from Table \ref{tab:geometry} into Eqn. \ref{p_general}, I get Figure \ref{fig:wave_plots}.
The figure was produced assuming slope angles $\alpha=0.1$. For other slope angles the TYORP pressure will scale as $\alpha^2$. 
The maximal TYORP for regolith in Figure \ref{fig:wave_plots} is about 30 times smaller than the maximal TYORP for spherical boulders in Figure \ref{fig:wave_plots}.
Therefore, at first sight, it may seem that regolith is unimportant.
Still, one must bear in mind several reservations.
Firstly, non-cracked boulders have big thermal parameters, $\theta=10\rightarrow 100$, while for regolith $\theta\sim 1$ \citep{golubov12}.
This will substantially suppress TYORP for boulders, while leaving TYORP for regolith close to its maximum.
Secondly, a much bigger fraction of surface is covered by regolith than is by boulders.
Lastly, the typical slopes $\alpha$ for asteroids are utterly unknown, and altering $\alpha$ can drastically decrease or increase TYORP.
The length scales I am referring to are millimeters to centimeters, which is far below the resolution of even \textit{in situ} observations of asteroids.

Summing up, one should acknowledge that regolith may have a substantial contribution to TYORP, which deserves a more detailed analysis than this rough estimate. Still, such an analysis would greatly divert us from the main focus of this paper.

\section{Integration of TYORP over all boulder sizes\label{sec-multiple}}
\subsection{Derivation of the integral for the overall TYORP}

All the previous analysis was done for boulders of only one particular size.
Let us now integrate TYORP over all sizes of boulders on the asteroid,
thus generalizing the numeric results obtained by \cite{sevecek15} for 25143 Itokawa.
For the size distribution of boulders, I take a power law: it is simple, reasonably precise, and the most widely used distribution. 
The interested reader is referred to \cite{asteroids4} for an excellent review of the available measurements of the boulder size distribution for different asteroids.

Consider a power-law distribution of boulder sizes,
\begin{equation}
 \frac{\mathrm{d}N}{\mathrm{d}l}=N_0 l^{-\gamma}\,,
\label{power-law0}
\end{equation}
where $\mathrm{d}N$ is the number of boulders with sizes between $l$ and $l+\mathrm{d}l$ lying on the entire surface area of the asteroid $s_\mathrm{ast}$, while $N_0$ and $\gamma$ are constant.
For us it is more convenient to re-write this distribution not in terms of the number of boulders, but in terms of the area covered by them:
\begin{equation}
 \frac{\mathrm{d}n}{\mathrm{d}l}=n_0 l^{2-\gamma}\,.
\label{power-law}
\end{equation}
Here $\mathrm{d}n=\mathrm{d}S/S_\mathrm{ast}$ is the fraction of the surface area covered by boulders with sizes between $l$ and $l+\mathrm{d}l$, while $n_0=\frac{N_0 s_\mathrm{proj}}{s_\mathrm{ast}l^2}$ is a new constant. 
(Remember that for geometrically similar boulders of different sizes $s_\mathrm{proj}\propto l^2$.)

Of course, a power law in the form of Eqs. (\ref{power-law0}) or (\ref{power-law}) can only be an approximation.
This distribution would necessarily diverge for either small or big boulders, and would necessarily eventually start disagreeing with the data for both small and big boulders. Still, I assume that deviations from the power law happen only for such small and such big boulders
that their contribution to TYORP is negligible anyway.

Let us determine the total dimensionless TYORP drag $p_\mathrm{tot}$ experienced by the surface as the dimensional TYORP drag force divided by $(1-A)\Phi S/c$.
Then $p_\mathrm{tot}$ can be obtained from $p(l)$ by integrating it over fractions of the surface area $\mathrm{d}n$ covered by boulders of each size:
\begin{equation}
 p_\mathrm{tot}=\int p(l)\,\mathrm{d}n=\int_{0}^{\infty} n_0 p(l)l^{2-\gamma}\,\mathrm{d}l
\label{definition-of-p}
\end{equation}
Given asymptotics of Eqn. (\ref{p_general}) $p\propto l^2$ for $l\rightarrow 0$ and $p\propto l^{-5}$ for $l\rightarrow \infty$, this integral converges for $\gamma\in(-2;5)$. Typical power indices for small boulders indeed lie in this range \citep{asteroids4,sevecek15}. For $\gamma$ outside this interval, finiteness of the integral has to be provided by setting the minimal or the maximal size of boulders, starting from which the power-law size distribution breaks.

From Eqn. (\ref{definition-of-p}) one can see the problem: the rational expression for $p$ provided by Eqn. (\ref{p_general}) does not look frightening only up until the moment one tries to integrate it over $l$.
Having spent many pleasant hours doing contour integration in the complex plane, I must acknowledge that the final result is too complicated to be of any practical use despite it being a closed form algebraic expression.
Moreover, when written in terms of real variables, it is far too lengthy to be accommodated into this subsection.

To get a more practical result from Eqn. (\ref{definition-of-p}), $p$ should be transformed into a simpler form.
Still, one can not neglect either term in Eqn. (\ref{p_general}): $a$, $b$, $\tau_0$, $l$ and $\theta$ are all of the order of unity, and thus all comparable with each other.
Therefore, I choose not to simplify the equation, but to brute-forcedly interpolate it.
This will cause some loss of accuracy, but as we will see, this loss is not much bigger than the errors already caused by the previous simplifications.

\subsection{Approximate expression for TYORP}

In this subsection I do a detour to find the best fit for $p$ as a function of $l$ and $\theta$ as given by Eqn. (\ref{p_general}).
In Figures \ref{fig:wall_maps} and \ref{fig:sphere_maps} one can see the three-dimensional surface $p(l,\theta)$ from above, and notice that in the $\lg{l}-\lg{\theta}$ plane lines of constant $p$ are roughly elliptical and roughly similar to each other.
In Figures \ref{fig:wall_plots} and \ref{fig:sphere_plots} one can see that vertical cross-sections of this three-dimensional surface look similar to Gaussian functions of the same width.
These observations allow us to guess that a decent approximation to $p(l,\theta)$ will be given by a two-dimensional log-normal function,
\begin{equation}
 p=p_0\mathrm{e}^{A_l\ln{l}+A_{\theta}\ln{\theta}+\frac{1}{2}A_{ll}\ln^2{l}+A_{l\theta}\ln{l}\ln{\theta}+\frac{1}{2}A_{\theta\theta}\ln^2{\theta}}.
\label{log-normal}
\end{equation}

\begin{figure}
  \centering    
  \includegraphics[width=0.45\textwidth]{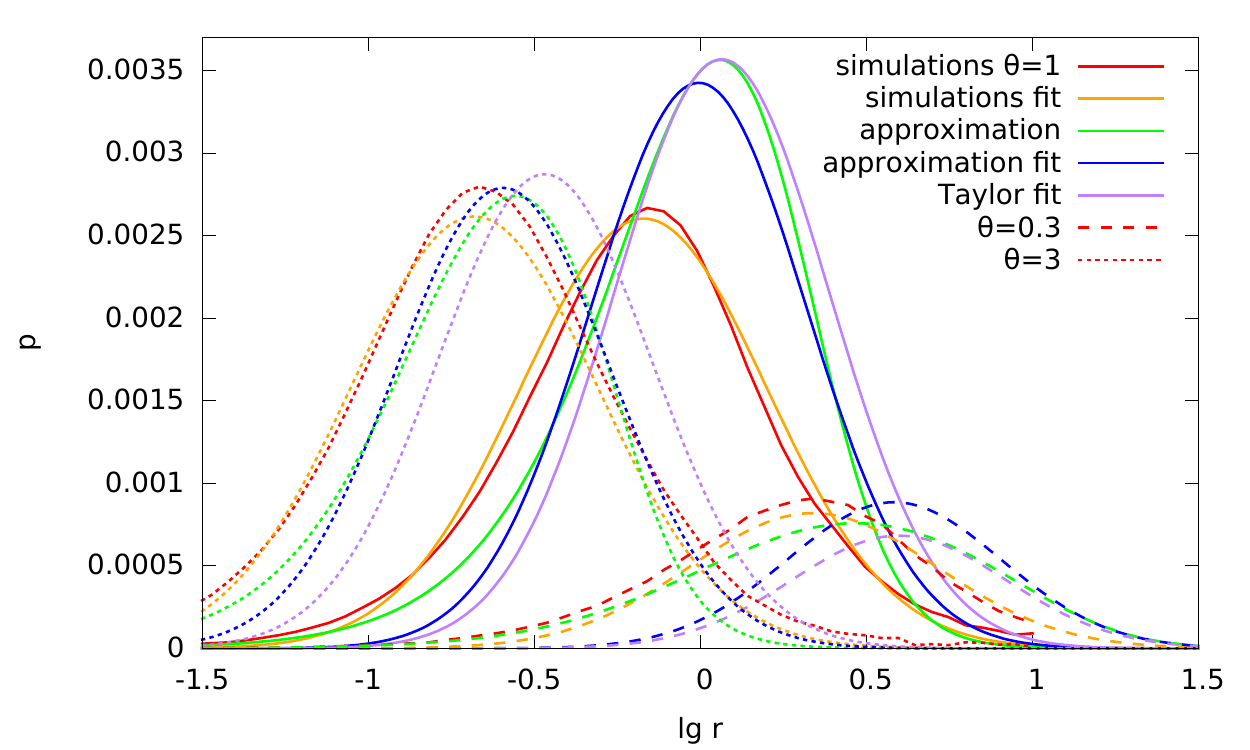}
  \caption{The comparison between different approximations for TYORP experienced by a spherical boulder. Three different types of lines (short dashed, long dashed and solid) show different values of the thermal parameter $\theta$. Different colors show different approximations: the exact numeric solution (\textit{red}), the log-normal function constructed by the least squares fit to the numeric solution (\textit{orange}), the approximate solution given by Eqn. \ref{p_general} (\textit{green}), the log-normal function constructed by the least squares fit to the approximate solution (\textit{blue}), and the log-normal function constructed by the Taylor decomposition of the approximate solution (\textit{purple}).}
  \label{fig:sphere_fit}
\end{figure}

There are several ways to choose the coefficients in this fitting function, and in Figure \ref{fig:sphere_fit} I explore different ways of fitting.
The least squares fit to the numeric solution is presented with the orange lines,
the least squares fit to the approximate solution Eqn. \ref{p_general} is plotted in blue.
Both fits are constructed for the same range of $r$ and $\theta$ as presented in Figure \ref{fig:sphere_maps}.
The purple lines are constructed by requiring Eqs. (\ref{p_general}) and (\ref{log-normal}) to have the same values in the point $l=\theta=1$, as well as the same first and second partial derivatives. For brevity, I call this approach the \textit{Taylor fit}, as I have indeed requested equality of the zeroth, first and second order terms in the Taylor series for $p(l,\theta)$ and its fit.
The analytically computed coefficients of the Taylor fit are presented in Table \ref{tab:taylor}.
Although cumbersome, they can still be instrumental to estimations of TYORP for complex-shaped bodies.

The exact numeric solution is overplotted with a red line, and the approximate solution given by Eqn. \ref{p_general} with a green line.
One can see that the least squares fit to the numeric result agrees with them very well, as do the two other fits agree with the approximate solution Eqn. \ref{p_general}. The Taylor fit to Eqn. \ref{p_general} is by construction perfect in the vicinity of the point $l=\theta=1$, but far away from this point it also works quite well.

Coefficients of the three considered fittings are given in Table \ref{tab:fitting}.

As a minor side result, in the same table I present $\alpha_0=\tan\left(\frac{1}{2}\arctan\frac{2A_{l\theta}}{A_{ll}-A_{\theta\theta}}\right)$, which determines orientation of the ellipsis with constant $p$ for the log-normal distribution. In the $\lg{l}-\lg{\theta}$ plane decrease of $p$ is slowest along the line $\theta\propto l^{\alpha_0}$. We see, that $\alpha_0$ obtained from different models are consistent with each other, and roughly consistent with the elongation of the red and yellow areas in Figure \ref{fig:sphere_maps}.

Now, having obtained the coefficients of the log-normal fit, I am ready to finish computing the integral for the overall TYORP.

\begin{table*}
\centering
\caption{Fitting coefficients for spherical boulders in different models\footnote{Coefficients $\mu$, $\nu$ and $\mathrm{ln}\theta_0$ are explained below in Section \ref{sec:integral}.}}
\begin{tabular}{l|cccccc|ccc|c}
Model & $p_0$ & $A_l$ & $A_\theta$ & $A_{ll}$ & $A_{\theta\theta}$ & $A_{l\theta}$ & $\mu$ & $\nu$ & $\mathrm{ln}\theta_0$ & $\alpha_0$ \\ 
\hline
Taylor fit to analytics & 0.00350 & 0.270 & 0.912 & -1.81 & -3.47 & -1.93 & 0.00763 & 1.191 & 0.442 & -0.659 \\
Least squares fit to analytics & 0.00342 & -0.020 & 0.475 & -1.81 & -3.69 & -2.13 & 0.00709 & 1.304 & 0.424 & -0.653 \\
Least squares fit to numerics & 0.00234 & -0.540 & -0.047 & -1.37 & -2.30 & -1.40 & 0.00644 & 1.518 & 0.580 & -0.723 \\ 
\end{tabular}
\label{tab:fitting}
\end{table*}

\begin{table}
\centering
\caption{Fitting coefficients in the Taylor fit}
\begin{tabular}{c|l}
Coefficient & Expression \\ 
\hline
$p_0$ & $\frac{8nab^2\tau_0^2CS}{(b+2\tau_0^3)(16\tau_0^6+a^2)(4(b+2\tau_0^3)^2+a^2)}$ \\ 
$A_l$ & $2-\frac{2\tau_0^3}{b+2\tau_0^3}-\frac{2a^2}{16\tau_0^6+a^2}-\frac{16(b+2\tau_0^3)\tau_0^3+4a^2}{4(b+2\tau_0^3)^2+a^2}$ \\ 
$A_\theta$ & $2-\frac{4a^2}{16\tau_0^6+a^2}-\frac{4a^2}{4(b+2\tau_0^3)^2+a^2}$ \\ 
$A_{ll}$ & $-\frac{2b\tau_0^3}{(b+2\tau_0^3)^2}-\frac{64a^2\tau_0^6}{(16\tau_0^6+a^2)^2}$ \\
& $-\frac{16\left(4a^2b^2+4b^3\tau_0^3+9a^2b\tau_0^3+16b^2\tau_0^6+4a^2\tau_0^6+16b\tau_0^9\right)}{(4(b+2\tau_0^3)^2+a^2)^2}$ \\ 
$A_{l\theta}$ & $-\frac{128a^2\tau_0^6}{(16\tau_0^6+a^2)^2}-\frac{64a^2(b+2\tau_0^3)(b+\tau_0^3)}{(4(b+2\tau_0^3)^2+a^2)^2}$ \\ 
$A_{\theta\theta}$ & $-\frac{256a^2\tau_0^6}{(16\tau_0^6+a^2)^2}-\frac{64a^2(b+2\tau_0^3)^2}{(4(b+2\tau_0^3)^2+a^2)^2}$ \\ 
\end{tabular}
\label{tab:taylor}
\end{table}

\begin{figure}
  \centering    
  \includegraphics[width=0.45\textwidth]{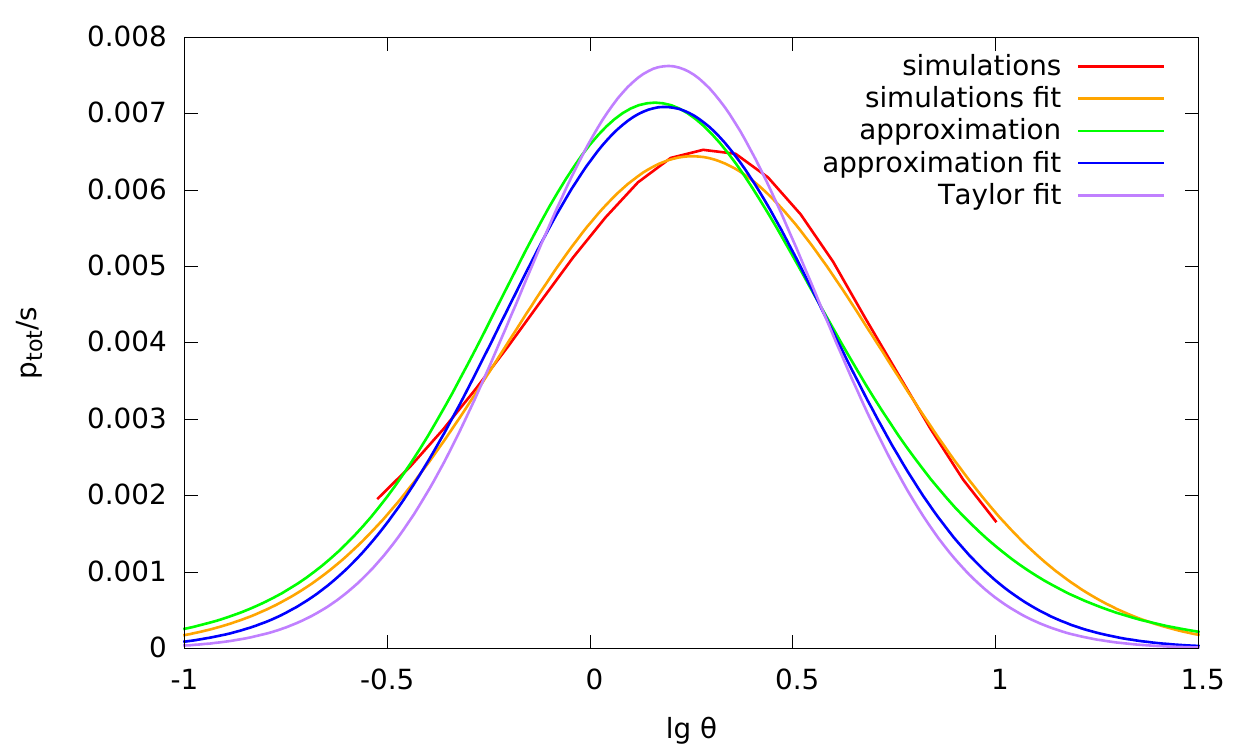}
  \caption{The normalized dimensionless pressure for an assembly of boulders with $\gamma=3$.
Different colors show different approximations: the exact numeric solution (\textit{red}), the log-normal function constructed by the least squares fit to the numeric solution (\textit{orange}), the approximate solution given by Eqn. \ref{p_general} (\textit{green}), the log-normal function constructed by the least squares fit to the approximate solution (\textit{blue}), and the log-normal function constructed by the Taylor decomposition of the approximate solution (\textit{purple}). As the exact solution was calculated only for $r\in[0.03;10]$, $\theta\in[0.3;10]$, points with $\theta$ outside the interval [0.3;10] were not plotted. On the other hand, $r$ outside the interval [0.03;10] were not discarded, but approximated using the least squares fit to the numeric solution, to make the integration more precize.}
  \label{fig:sphere_integral}
\end{figure}

\subsection{Computing the integral for the overall TYORP}
\label{sec:integral}

Returning to the main focus of this section,
I still need to integrate the simplified expression for TYORP over all boulder sizes.
Substituting Eqn. (\ref{log-normal}) into Eqn. (\ref{definition-of-p}) and computing the Gaussian integral, I get
\begin{equation}
 p_\mathrm{tot}=n_0\mu\mathrm{e}^{-\frac{\left(\ln{\theta}-ln{\theta_0}\right)^2}{\nu^2}},
\label{log-normal-integral}
\end{equation}
where three new constants have been introduced:
\begin{eqnarray}
\label{log-normal-coef}
\mu&=&p_0\sqrt{\frac{2\pi}{-A_{ll}}}\mathrm{e}^{\frac{2A_\theta A_{l\theta}(A_l-\gamma+3)-A_{\theta\theta}(A_l-\gamma+3)^2-A_{\theta}^2A_{ll}}{2(A_{\theta\theta}A_{ll}-A_{l\theta}^2)}}, \\ \nonumber
\nu&=&\sqrt{\frac{-2A_{ll}}{A_{\theta\theta}A_{ll}-A_{l\theta}^2}}, \\ \nonumber
\ln\theta_0&=&\frac{A_{l\theta}(A_l-\gamma+3)-A_\theta A_{ll}}{A_{\theta\theta}A_{ll}-A_{l\theta}^2}. 
\end{eqnarray}

The mathematical expressions are again somewhat lengthy, but they are worth the agreement between the different estimates of $p_\mathrm{tot}$ we see in Figure \ref{fig:sphere_integral}.
Even the substantial discrepancies between the numeric and the analytic solutions observed in Figure \ref{fig:sphere_fit} are mostly eliminated by integration in Figure \ref{fig:sphere_integral}.
The resulting discrepancies between the exact results and the different approximations lie within about 20\% of the maximal value on the plot,
thus certifying that both of our approximations are good for estimating the TYORP effect.
The constants Eqn. (\ref{log-normal-coef}) are also listed in Table \ref{tab:fitting}, and one can see that they are close to each other.

I now apply these results to asteroid 25 143 Itokawa, both to estimate the TYORP and to appease myself by certifying the uncertainties in physical properties of even such a well-studied asteroid cause a much bigger error in the TYORP acceleration than the simplifications underlying my analytic model.
The assumed physical properties of boulders taken from \cite{sevecek15} are listed in Table \ref{tab:itokawa}.
Following \cite{sevecek15}, I use two different values of the heat conductivity of rock: bigger $\kappa_1$ presumably corresponds to solid rock, while smaller $\kappa_2$ could correspond to cracked rock, which is assumed to be widely present on atmosphereless bodies \citep{delbo14}.
Respectively, I get two different thermal parameters $\theta_1=13$ and $\theta_2=4.1$.
I use the size distribution of boulders from \cite{sevecek15}, $dN=(14\pm 9)\cdot 10^3 L^{-3.0\pm 0.2}dL$, where $dL$ is in meters.
For simplicity, I choose the power index $\gamma=3$. In this case $n_0$ is independent of $L_\mathrm{cond}$. 
Nondimensionalizing $L$, transforming from the number of boulders $dN$ into their relative surface area $dn$, assuming roughly circular projections of boulders, and taking for the surface area of Itokawa $3.93\cdot 10^5$ m$^2$ \citep{demura06}, I get $n_0=0.028\pm 0.018$.
From Figure \ref{fig:sphere_integral}, I find corresponding pressures $P(\theta_1)\approx 0.001$ and $P(\theta_2)\approx 0.005$.
Estimating the dimensionless TYORP of the whole asteroid as $\tau_z\approx 9sT(\theta)$ \citep{golubov14}, I get for solid rocks $\tau_{z1}=0.00025\pm 0.00015$ and for cracked rocks $\tau_{z2}=0.0013\pm 0.0007$.
This value should be compared to the difference $\Delta\tau_z= 0.003\pm 0.002$ \citep{golubov14} between the observed accelerating torque acting on Itokawa and the predicted normal YORP torque, as one may expect $\Delta\tau_z$ to be equal to Itokawa's TYORP. One can see that the cracked rock gives a value consistent with $\Delta\tau_z$, while the solid rock gives a value an order of magnitude smaller than $\Delta\tau_z$.
As a side remark, I note that the tens-of-per-cent uncertainty of the theoretical model is much smaller than an order of magnitude uncertainty caused by the uncertainty of the physical properties of the asteroid.
The most important takeaway message from this estimate for 25 143 Itokawa is that for realistic properties of boulders and their realistic shape distribution, Eqn. (\ref{log-normal-integral}) predicts a significant effect, comparable to the normal YORP.

\begin{table}
\centering
\caption{Parameters used in the TYORP estimation for 25 143 Itokawa}
\begin{tabular}{l|c|l}
Physical quantity & Notation & Value \\ 
\hline
Heat conductivity & $\kappa_1$ & 2.65 W m$^{-1}$ K$^{-1}$ \\
& $\kappa_2$ & 0.26 W m$^{-1}$ K$^{-1}$ \\
Specific heat capacity & $C$ & 680 J kg$^{-1}$ K$^{-1}$ \\
Density & $\rho$ & 2700 kg m$^{-3}$ \\
Albedo & $A$ & 0.1 \\
Emissivity & $\epsilon$ & 0.9 \\
Angular velocity & $\omega$ & $\frac{2\pi}{12.132\,\mathrm{h}}$ \\
Solar constant & $\Phi$ & $\frac{1361}{1.324^2}$ W m$^{-2}$ \\
\end{tabular}
\label{tab:itokawa}
\end{table}

\section{Discussion\label{sec-discussion}}

Reviewing the derivation of the analytic expression of TYORP (Eqn. (\ref{p_general})), one can single out three major ingredients sufficient for TYORP to appear:

1. Heating pattern with a day--night asymmetry ($C\neq 0$) and an east--west asymmetry ($S\neq 0$).

2. Thermal inertia ($0<\theta<\infty$, $0<a<\infty$) and thermal conduction ($0<b<\infty$).

3. Non-linearity of the heat emission law.

The non-linearity enters the equations for TYORP in two related ways: through the boundary condition of the heat conduction equation (Eqn. (\ref{heat_balance})) and through the definition of TYORP pressure (Eqn. (\ref{p_def})).
The former's contribution is negative, the latter's contribution is positive, and the latter wins, as can be seen from the substitution of Eqn. (\ref{2nd_2}) into Eqn. (\ref{p_series}).

Acknowledging the generality of conditions 1--3, one must conclude that TYORP is a very general feature, which is not bound to any particular boulder shape or illumination pattern. More or less any boulder shapes can be substituted into Eqn. (\ref{heat_balance1}), while Eqn. (\ref{insolation}) presents a most generic illumination function, which is still capable of causing TYORP of a realistic magnitude, although boulders with steeper slopes (bigger $n$) and illuminated in a more asymmetric way (bigger $S$) are subject to systematically bigger TYORP, as can be seen from Eqn. (\ref{p_general}).

In my model, $C$ is always positive, as the boulder is illuminated in the day, not at night.
$S$ is usually positive, as all realistic boulders are more illuminated from the east before noon,
and more illuminated from the west after noon.
Thus all terms in Eqn. (\ref{p_general}) are positive, and the resulting TYORP derived from the analytic model should accelerate the rotation of the asteroid under any realistic conditions, not decelerate it.
This fact has already been observed for different geometries of boulders by \cite{golubov14} and \cite{sevecek15} in their numerical simulations.

This holds not only for convex structures like boulders or mounds, but also for concave structures, like grooves or pits on the surface of the asteroid.
In the latter case the mechanism is the same as usual: east-facing slopes are better illuminated in the morning, west-facing slopes in the afternoon, and the heat conduction occurs between them. 
The problem is again approximately described by Eqs. (\ref{heat_balance1}) and (\ref{insolation}) with all constants being positive, and the solution again leads to Eqn. (\ref{p_general}).
It means that the west-facing slopes again emit on average more light, push the surface more in the eastern direction, and again accelerate the rotation of the asteroid.

Thus most imaginable surface structures should be to a certain extent subject to positive TYORP: crater rims, crater pits, grooves and cracks on the surface, small mounds and summits, or just wavy patterns of regolith, rough surfaces of stones, lone boulders and boulder fields... A heap of pebbles can contribute to TYORP either as a single body, or via individual pebbles, or via individual pocks and wedges of individual pebbles, and which of these descriptions is the most appropriate depends on the rotation rate of the asteroid, the heat conductivity of rock, as well as other parameters. Singling out the most important contributors to TYORP and accounting for them all is an important task for future research.

\section{Conclusions\label{sec-conclusions}}

I have constructed an approximate fully analytic model of TYORP, in which I used an averaging procedure to transform a partial differential equation into an ordinary differential equation, then used perturbation theory to transform the latter into algebraic equations, and finally arrived at an algebraic expression for TYORP.
Along the way the shape of the boulder was boiled down to just a few geometric coefficients, and the illumination pattern -- to its few Fourier harmonics.
Despite the simplicity, the analytic expression for TYORP was found to be in good agreement with numeric simulations for a stone wall \citep{golubov12} and a spherical boulder \citep{golubov14}.

I used the analytic model to estimate the TYORP produced by the non-smoothness of regolith on an asteroid.
On the one hand, regolith should be much flatter than boulders, which diminishes its TYORP.
On the other hand, it has a more favorable thermal parameter $\theta$ and presumably covers a bigger fraction of the surface, which increases its TYORP contribution.
Given the utter uncertainty of the unevenness of regolith on different scales, it is hard to say whether or not its contribution is significant, but neglecting it now seems unsafe.

I then have integrated TYORP over boulders of all different sizes, assuming a power-law size distribution for boulders, and doing a two-dimensional log-normal fit to the analytic TYORP drag.
The additional error introduced by this fit was generally not bigger than the error already caused by the assumptions underlying the analytical model.
Moreover, the error substantially diminished after integration over the boulder sizes.
The resulting TYORP drag of the entire assembly of boulders appeared log-normal in terms of the thermal parameter $\theta$, and concurrently in terms of rotation rate of the asteroid.
It means that even for a broad distribution of boulders over sizes, TYORP still has a relatively narrow peak, a corresponding rotation rate with the biggest TYORP, and a relatively fast decrease of TYORP for faster or slower rotation.
The integrated TYORP drags for the log-normal approximation, for the full analytic solution and for the numeric simulation were found to be in good agreement with each other.

The proposed analytic model demonstrates that TYORP should appear for most realistic shapes of boulders or regolith, and should	 be positive in most realistic cases.

\section*{Acknowledgements}
I am very grateful to Daniel J. Scheeres for discussing with me the astronomical content of the problem, to Uliana Pyrohova for counseling me with mathematical issues, to Vlad Unukovych for reviewing the physics of the paper, and to Alexander Kostenko for improving the language style.
Without the help of either of them the paper would have been stuck at a certain point.

\label{lastpage}


\begin{thebibliography}{99}
\bibitem[\protect\citeauthoryear{Delbo et al.}{2014}]{delbo14}
Delbo M., Libourel G., Wilkerson J., et al., 2014, Nature 508, 233 
\bibitem[\protect\citeauthoryear{Demura et al.}{2006}]{demura06}
Demura H., Kobayashi S., Nemoto E., et al., 2006, Science 312, 1347
\bibitem[\protect\citeauthoryear{Golubov \& Krugly}{2011}]{golubov11}
Golubov O., Krugly Yu.~N., 2011, Planetary Defence Conference, From Threat to Action, held 09-12 May 2011, in Bucharest, Romania 
\bibitem[\protect\citeauthoryear{Golubov \& Krugly}{2012}]{golubov12}
Golubov O., Krugly Yu. N., Tangential component of the YORP effect, 2012, ApJL 752, L11
\bibitem[\protect\citeauthoryear{Golubov et al.}{2014}]{golubov14}
Golubov O., Scheeres D. J., Krugly Yu. N., A three-dimensional model of tangential YORP, 2014, ApJ 794, 22
\bibitem[\protect\citeauthoryear{Murdoch et al.}{2015}]{asteroids4}
Murdoch N., S{\'a}nchez P., Schwartz S.~R., Miyamoto H., 2015, Asteroids IV, 767
\bibitem[\protect\citeauthoryear{\v{S}eve\v{c}ek et al.}{2015}]{sevecek15}
\v{S}eve\v{c}ek P., Bro{\v z} M., {\v C}apek D., \& {\v D}urech, J., The thermal emission from boulders on (25143) Itokawa and general implications for the YORP effect, 2015, MNRAS 450, 2104
\bibitem[\protect\citeauthoryear{\v{S}eve\v{c}ek et al.}{2016}]{sevecek16}
\v{S}eve\v{c}ek, Golubov O., Scheeres D.J., Krugly Yu. N., Obliquity dependence of the tangential YORP, 2016, A\&A 592, 115
\end{thebibliography}
\end{document}